\pdfoutput=1
\documentclass[11pt]{article}

\usepackage{acl}

\usepackage{times}
\usepackage{latexsym}
\usepackage{booktabs}
\usepackage{enumitem}
\usepackage{multirow}
\usepackage{amsmath}
\usepackage{xspace}
\usepackage{graphicx}
\usepackage[T1]{fontenc}
\usepackage[utf8]{inputenc}

\usepackage{microtype}

\usepackage{inconsolata}
\usepackage{wasysym}

\usepackage{graphicx}

\newcommand{\methodname}{\textsc{MythTriage}\xspace}

\makeatletter
\def\@fnsymbol#1{\ensuremath{\ifcase#1\or \dagger\or \ddagger\or
\mathsection\or \mathparagraph\or \|\or **\or \dagger\dagger
\or \ddagger\ddagger \else\@ctrerr\fi}}
\makeatother
\usepackage{xurl}

\newcommand\edit[1]{{\textcolor{black}{#1}}}

\title{\methodname: Scalable Detection of Opioid Use\\Disorder Myths on a Video-Sharing Platform}

\author{Hayoung Jung\textsuperscript{1}\thanks{Most work done at the University of Washington.} \ \ \ \ \ \ \ Shravika Mittal\textsuperscript{3} \ \ \ \ \ \ \ Ananya Aatreya\textsuperscript{2} \\
\textbf{Navreet Kaur\textsuperscript{2}} \ \ \ \textbf{Munmun De Choudhury\textsuperscript{3}} \ \ \ \textbf{Tanushree Mitra\textsuperscript{2}} \\
\textsuperscript{1}Princeton University \\  
\textsuperscript{2}University of Washington \\ 
\textsuperscript{3}Georgia Institute of Technology \\
\small{\textbf{Correspondence: }\texttt{hayoung@cs.princeton.edu}, \texttt{tmitra@uw.edu}}
}

\begin{document}
\maketitle
\begin{abstract}

Understanding the prevalence of misinformation in health topics online can inform public health policies and interventions. However, measuring such misinformation at \textit{scale} remains a challenge, particularly for high-stakes but understudied topics like opioid-use disorder (OUD)---a leading cause of death in the U.S. We present the first large-scale study of OUD-related myths on YouTube, a widely-used platform for health information. With clinical experts, we validate 8 pervasive myths and release an expert-labeled video dataset. To scale labeling, we introduce \methodname, an efficient triage pipeline that uses a lightweight model for routine cases and defers harder ones to a high-performing, but costlier, large language model (LLM). \methodname achieves up to 0.86 macro F1-score while estimated to reduce annotation time and financial cost by over 76\% compared to experts and full LLM labeling. We analyze 2.9K search results and 343K recommendations, uncovering how myths persist on YouTube and offering actionable insights for public health and platform moderation.\vspace{-0.5mm}\footnote{Code and Data: \url{https://github.com/hayoungjungg/MythTriage}, Models: \url{https://huggingface.co/SocialCompUW/youtube-opioid-myth-detect-M1}}

\textcolor{red}{\textit{Warning: Some content of this paper, included to contextualize our data, is misleading.}}\vspace{-2.4mm}
\end{abstract}

\begin{figure*}[t!]
    \centering
    \includegraphics[width=\textwidth]{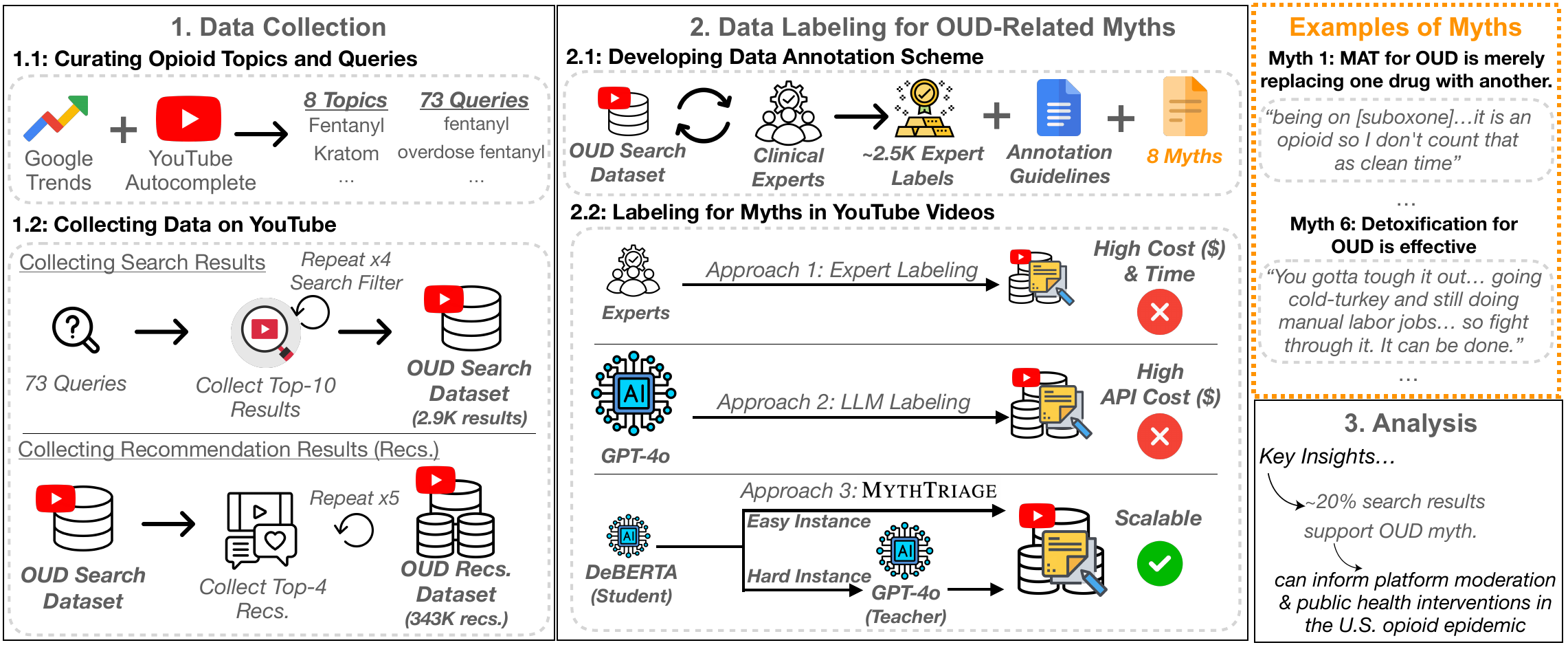}\vspace{-3mm}
    \caption{
    \textbf{Study Overview.} (1.) We curated opioid-related topics and queries (1.1), then collected YouTube search and recommendation results (1.2). (2.) To label myths, clinical experts validated 8 myths (with examples shown in the orange box), refined the annotation guidelines, and provided 2.5K labels (2.1). We compare three potential labeling approaches: expert labeling, LLM labeling, and \methodname---a scalable pipeline using lightweight distilled models for easy cases and defers hard ones to high-performing, but costly LLM. (3.) Using \methodname, we analyzed the labeled dataset at \textit{scale}, offering actionable insights for platform moderation and public health.
    }
    \label{fig:main}\vspace{-2mm}
\end{figure*}

\section{Introduction}

Online platforms are a key source of health information \cite{dubey2014analysis}, with video-sharing platforms like YouTube playing an increasingly prominent role in shaping public understanding during public health crises \cite{bora2018internet, KHATRI2020101636}. 
However, online platforms are also a conduit for widespread misinformation that can undermine public health efforts \cite{factchecker}. A particular instance is the case of opioid use disorder (OUD) -- a leading cause of death in the U.S, with 108K drug overdose deaths in 2022 \cite{nihDrugOverdose}. Facing offline stigma, individuals with OUD often rely on online platforms for health information and recovery guidance \cite{Balsamo_Bajardi_De_Francisci_Morales_Monti_Schifanella_2023}. But online myths---e.g., \textit{medication for addiction treatment (MAT) is simply replacing one drug with another}---fuel treatment hesitancy, distrust in healthcare, and stigma \cite{info:doi/10.2196/30753, doi:10.1177/1178221816685087}. This has undermined efforts to promote clinically-approved MAT \cite{nationalacademiesMedicationsOpioid}. 

Understanding the scale and spread of such misinformation is crucial for health officials and platforms seeking to design effective interventions \cite{10.1145/3411764.3445250}. While prior works have acknowledged this gap and explored social dynamics and discourse in online health communities \cite{bunting2021socially, chancellor2019discovering}, large-scale analyses of the OUD-related myth prevalence, especially on video-sharing platforms, remain limited. 
Detecting misinformation on video platforms \textit{at scale} remains challenging, as it requires domain expertise and intensive labeling of large volumes of content. While recent works highlight the potential of large language models (LLMs) to address this scale challenge in social science research \cite{ranjit-etal-2024-oath, dammu-etal-2024-uncultured}, their increasing compute demands and high API inference cost---especially on long-form video content---limit their widespread use for large-scale misinformation detection, particularly in high-stakes health issues. 

% method
To address these gaps, we present the first large-scale study of OUD-related myths on YouTube, illustrated in Figure \ref{fig:main}. We construct two datasets: \texttt{OUD Search Dataset} of 2.9K search results (1.8K unique videos) from 73 trending queries across four opioid and four treatment topics, and \texttt{OUD Recommendation Dataset} of 343K recommendations (164K unique videos) obtained by crawling the top-four recommendations per unique video in \texttt{OUD Search Dataset}, going five levels deep. In collaboration with clinical experts, we validate 8 pervasive myths (see Table \ref{tab:oud-myths} for the list of myths and examples), refine the annotation guidelines, and construct a gold-standard dataset of 310 videos labeled across all myths, totaling 2.5K expert labels. 

To scale beyond expert or full-LLM labeling, we introduce \textbf{\methodname}, an efficient triage pipeline inspired by model cascade architectures \cite{varshney-baral-2022-model, mamou2022tangobertreducinginferencecost} (see Figure \ref{fig:main}). \methodname uses a lightweight model for routine cases and defers harder ones to a high-performing, but costly LLM. We evaluate ten open-weight and proprietary LLMs (see Table \ref{tab:model-performance-myths-f1-score}) on our gold-standard dataset and distill a strong lightweight model from \texttt{GPT-4o}. \methodname achieves strong performance across myths (0.68-0.86 macro F1-scores; median 0.81), while estimated to reduce the annotation cost by 98\% and time by 96\% compared to expert labeling---and achieving 94\% cost and 76\% time savings over full LLM labeling of the \texttt{OUD Recommendation Dataset}. \methodname offers scalable, cost-effective detection of OUD myths across large video corpora, facilitating large-scale analysis.

Using the annotated labels, we offer the first large-scale empirical view into OUD-related myth prevalence on YouTube. Overall, nearly 20\% of the search results support myths. Notably, videos related to Kratom, a widely-used drug falsely promoted as a non-addictive and safe alternative to opioids \cite{mayo-clinic}, contained more myth-supporting content (36\%) than those opposing (22\%). We find that 12.7\% of recommendations to myth-supporting videos lead to other supporting videos at the initial recommendation level, rising to 22\% at deeper levels. These findings reveal the scale and persistence of OUD-related myths on the platform. Our results offer actionable insights for public health and platform moderation, demonstrating the value of \methodname and highlighting opportunities for intervention in the context of an ongoing and high-stakes opioid crisis.

\section{Related Works}

\noindent\textbf{LLMs for Social Science and Health 
Applications.} LLMs have been increasingly used in social science and health research \cite{park-etal-2024-valuescope, sharma-etal-2023-cognitive, li2024mediq}, particularly for data annotation tasks \cite{ranjit-etal-2024-oath, dammu-etal-2024-uncultured, ziems-etal-2024-large, tan-etal-2024-large}. However, their high compute demands and API costs limit their scalability for large-scale annotation tasks \cite{ding2024unleashingreasoningcapabilityllms}. To reduce costs, prior work has proposed model cascading frameworks that combine lightweight models with stronger models for uncertain predictions \cite{khalili2022babybearcheapinferencetriage, varshney-baral-2022-model, geifman2017selectiveclassificationdeepneural, 10.1007/978-3-030-30484-3_26, mamou2022tangobertreducinginferencecost}. Yet, few have integrated LLMs into these cascades to address the scalability challenge \cite{farr2024llmchainensemblesscalable}, particularly for practical and high-stakes applications like large-scale misinformation detection for OUD. 

\edit{In contrast to prior model cascading frameworks evaluated mainly on standard benchmarks such as CoLA \cite{varshney-baral-2022-model}, SQuAD \cite{mamou2022tangobertreducinginferencecost}, or CIFAR-10 \cite{10.1007/978-3-030-30484-3_26}, our work demonstrates the practical value of these frameworks by integrating and validating them in a real-world, high-stakes health contexts, contributing a scalable, extensible pipeline for large-scale OUD myth detection in video-sharing platforms.}

\noindent\textbf{Stigma and Misinformation in High-Stakes Health Contexts.} Prior works have investigated online platforms and LLMs for stigma and misinformation in high-stakes health contexts \cite{jung2025algorithmicbehaviorsregionsgeolocation, nguyen2024supportersskepticsllmbasedanalysis, kaur-etal-2024-evaluating}, with efforts to employ LLMs to reduce stigma and support well-being \cite{song2025typingcureexperienceslarge, mittal2025exposurecontentwrittenlarge}. Among health conditions, OUD is among the most stigmatized, often viewed as a result of ``willful choice'' rather than a chronic, treatable disease \cite{corrigan2017making, kennedy2017social}. Such a narrative fuels persistent myths that undermine harm reduction and MAT \cite{doi:10.1080/10826084.2022.2079133, KRUIS2021108485}. To escape offline stigma, many turn to online recovery communities for support and information \cite{DAGOSTINO20175, Balsamo_Bajardi_De_Francisci_Morales_Monti_Schifanella_2023}, but these spaces also contain harmful OUD myths. A few studies have quantified OUD-related myths online \cite{oud-audit, info:doi/10.2196/30753, info:doi/10.2196/44726}, but these efforts have been limited to a small set of myths and text-based platforms like Reddit and Twitter. Building on these efforts, we collaborate with clinical experts and introduce \methodname, a scalable pipeline for detecting 8 distinct OUD-related myths across large video corpora. This work presents the first large-scale empirical analysis of OUD-related myth prevalence on YouTube, a challenging task that requires collaboration with clinical experts.

\section{Data Collection}\label{sec:data-collection}
To collect OUD-related data on YouTube, we outline a two-step process: 1) curating OUD-related topics and associated search queries, and 2) performing large-scale data collection on the platform.

%%% full list of search queries
\begin{table}[t]
\scriptsize
\centering
\begin{tabular}{p{2.5cm}p{3.2cm}}
\toprule
\textbf{Opioid Topics} &
  \textbf{Sample Search Queries} \\\midrule
  Fentanyl & fentanyl, overdose fentanyl
  \\\cline{2-2}
Percocet (Oxycodone) & percocet, oxycodone
  \\\cline{2-2}
Heroin & heroin addict, on heroin \\\cline{2-2}
Codeine &
codeine, codeina
  \\\cline{2-2}
Kratom &
kratom withdrawal, kratom
  \\\cline{2-2}
Narcan & 
narcan, narcan training
  \\\cline{2-2}
Suboxone &
suboxone, suboxone withdrawal
  \\\cline{2-2}
Methadone&
methadone, methadone clinic\\
\bottomrule
\end{tabular}
\caption{For each topic, we provide a sample of our curated search queries. The top four are opioid-related topics, and the bottom four are MAT-related. See Table \ref{tab:query-set} for the full 73 queries.}\vspace{-3mm}
\label{tab:sample-query}
\end{table}

\subsection{Curating Opioid Topics and Queries}\label{sec:topics}

\subsubsection{Selecting Topics}\label{sec:selecting-topics}

To identify opioid topics, we used a curated lexicon of 156 keywords covering opioid drugs, medication-assisted treatments (MAT), and prescription medicines. This lexicon---developed in consultation with public health experts and prior literature in \citet{info:doi/10.2196/30753}---includes generic names (e.g., Oxycodone), trade names (e.g., OxyContin), and street names (e.g., OC) to ensure comprehensive coverage of opioid-related topics.

Since collecting data for all 156 keywords is impractical, we used Google Trends (\textit{Trends}) to identify the four most popular opioid and four most popular MAT keywords, yielding eight keywords in total. \textit{Trends} reflects real-world search interests and suggests related queries. We systematically filtered out overly broad keywords (e.g., ``Water''), those lacking \textit{Trends} data, or those with fewer than five related queries, filtering the set to 28 keywords (see Table \ref{tab:opioid-popularity}). \edit{The complete set of the 28 keywords is shown in Appendix Table~\ref{tab:opioid-popularity}.}

\edit{Using the set of 28 keywords}, we conducted pairwise comparisons of keywords in \textit{Trends} to rank them by popularity.\footnote{\textit{Trends} limits comparisons to five keywords and lacks an API, so we used \citet{serpAPI} to automate the comparisons.} For each pair, \textit{Trends} returns relative interest scores (ranging from 0 to 100); the higher-scoring keyword is considered more popular and thus ``wins.'' We computed win-rates across all pairs (see Table \ref{tab:opioid-popularity}). To identify the most popular topics across opioids and MAT, we selected the top 4 opioid and MAT keywords,\footnote{While not MAT, we include ``Kratom'' due to debunked claims of its efficacy in treating OUD \cite{mayo-clinic}.} with the highest win-rates \edit{from the set of 28 keywords}, yielding 8 final topics (Table \ref{tab:sample-query}). \S \ref{appendix:search-topic} details \textit{Trends} configuration and pairwise comparison.

\subsubsection{Selecting Search Queries}\label{sec:search-queries} 

To capture how users search for each topic on YouTube, we used \textit{Trends} and YouTube autocomplete suggestions to identify representative queries \cite{10.1145/3392854}. Since \textit{Trends} returns popular related queries on YouTube, we gathered related queries per topic (see \S \ref{appendix:trends-config} for details). To obtain additional trending queries, we also collected the top-10 autocomplete suggestions from YouTube search per topic. In total, this yielded 225 queries across 8 topics (see Table \ref{tab:sample-query} for sample queries).

To refine the query list, two researchers with prior experience in OUD-related myths qualitatively filtered queries following \citet{jung2025algorithmicbehaviorsregionsgeolocation}. In particular, we excluded queries that were overly broad (e.g., ``overdose''), overly specific (e.g., ``percocet future lyrics''), non-English queries, duplicates, mentioned individuals (e.g., ``lil wyte''), or fell outside the scope of OUD and MAT (e.g., ``fentanyl cop''). Similar queries were manually merged (e.g., ``the opioid crisis'' and ``opioid crisis''). This yielded 73 queries across 8 topics (see Table \ref{tab:query-set}), capturing trending and relevant search interests. 

\begin{table}[t!]
\centering
\resizebox{\columnwidth}{!}{%
\begin{tabular}{lcccc}
\toprule
\textbf{Dataset} & \textbf{Total} & \textbf{Unique} & \textbf{Expert} & \textbf{Synthetic} \\
\midrule
\texttt{OUD Search} & 2,893   & 1,776   & 310 & 1,466 \\
\texttt{OUD Recommendation} & 342,707 & 164,085 & 0   & 164,085 \\
\bottomrule
\end{tabular}}
\caption{Summary of the \texttt{OUD Search} and \texttt{OUD Recommendation Dataset}, showing the number of total videos (\textit{Total}), unique videos (\textit{Unique}), expert-annotated videos (\textit{Expert}), and \texttt{GPT-4o}-labeled videos (\textit{Synthetic}). \edit{Expert annotations were focused on the \texttt{OUD Search Dataset} to build a gold-standard dataset (\S\ref{sec:expert-annotation}). Both datasets require the same video labeling, but the \texttt{OUD Recommendation Dataset} is much larger (164K videos) and noisier, making expert annotation infeasible; we therefore used \methodname for efficient labeling.}}
\label{tab:oud-datasets}
\end{table}

\subsection{Collecting Data on YouTube}
With the topics and queries, we collected YouTube search and recommendation results to measure the prevalence of OUD-related myths. \edit{Table~\ref{tab:oud-datasets} summarizes the datasets, reporting counts of total, unique, expert-annotated, and \texttt{GPT-4o}-labeled videos.}

\subsubsection{Collecting Search Results}\label{sec:search-result-collection}
Next, we used the 73 curated queries (\S\ref{sec:search-queries}) to query and collect YouTube search results via the Data API \cite{youtube-api} to assess the prevalence of OUD myths. The API ranks content by query relevance and engagement metrics, without user data, ensuring results are not personalized \cite{youtube-data-api}. The API allows sorting the search results by four search filters: ``Relevance,'' ``Upload Date,'' ``View Count,'' and ``Rating.'' For each query and search filter, we collected the top 10 search results, as nearly 95\% of user traffic goes to the first page of the search results \cite{web_traffic}. 

For each video, we gathered metadata, including title, description, transcript, tags, and its rank in the results, creating the \texttt{OUD Search Dataset} with 2,893 search results (1,776 unique videos; \edit{see Table~\ref{tab:oud-datasets}).}

\subsubsection{Collecting Recommendation Results}\label{sec:recommendation-collection}

To measure the prevalence of myths in YouTube recommendations, we used a cascaded approach following \citet{Albadi_2022} to gather recommendations from December 18th to 20th, 2024. Using Google's InnerTube API \cite{inntertube}, we collected the top four recommended videos per unique video in the \texttt{OUD Search Dataset}, as in \cite{Albadi_2022}. This yielded 6,356 \textit{level 1} video recommendations (3,107 unique videos). We repeated this process through \textit{level 5}, collecting the top four recommendations per unique video at the previous level, resulting in a total of 342,707 recommendation links and 164,085 unique videos across all levels \edit{(Table~\ref{tab:oud-datasets}). Detailed counts of videos per recommendation levels are provided in Table~\ref{tab:recommendation_levels}.} We retrieved the metadata for these videos through the YouTube Data API to create the \texttt{OUD Recommendation Dataset}.\footnote{We could not retrieve data for 274 videos (0.2\% of data).}

\section{Developing Data Annotation Scheme}
To label videos for myths, we underwent extensive procedures to identify myths, develop the data annotation scheme, and create the expert-labeled gold standard dataset. For brevity, we detail the annotation scheme development, guidelines, and expert labeling process in \S\ref{appendix:annotation-appendix}.

\begin{table}[t!]
\centering
\scriptsize
\resizebox{\columnwidth}{!}{%
\begin{tabular}{@{}p{3.95cm} p{3.95cm}@{}}  
\toprule
\multicolumn{2}{c}{\textbf{OUD-Related Myths}} \\
\midrule

M1: MAT is merely replacing one drug with another. 
& M5: Physical dependence or tolerance is the same as addiction. \\

\addlinespace[0.5ex]

M2: OUD is a self-imposed condition, not a treatable disease. 
& M6: Detoxification for OUD is effective. \\

\addlinespace[0.5ex]

M3: The ultimate treatment goal for OUD is abstinence from any opioid use. 
& M7: You should only take medication for a brief period of time. \\

\addlinespace[0.5ex]

M4: Only patients with certain characteristics are vulnerable to addiction. 
& M8: Kratom is a non-addictive, safe alternative to opioids. \\

\bottomrule
\end{tabular}}
\caption{List of 8 OUD-related myths examined in our study. Some myths were paraphrased for brevity. See Table \ref{tab:oud-myths} for representative examples.}
\label{tab:oud-myths-singlecol}
\end{table}

% Data annotation scheme (how we developed it)
\subsection{How do we know what is a myth?}\label{sec:identify-myth}

To identify myths, we drew from prior literature and clinical sources. \citet{info:doi/10.2196/44726} conducted a systematic review of four online platforms, where three public health experts identified five prevalent OUD-related myths based on substance use literature \cite{PMID:30452633,doi:10.1056/NEJMp1802741}. We supplemented these with three additional myths from clinical sources, debunking pervasive myths about MAT \cite{suboxone-myths, buprenorphine-myths, suboxone-myth, kratom-myths}. All selected myths are recognized by major health organizations, such as the U.S. \citet{amcp} and \citet{johns-why-myths}, and were validated by clinical researchers, as described below. Table \ref{tab:oud-myths-singlecol} lists the 8 myths.

\subsection{Expert-Annotated Gold Standard Dataset}\label{sec:expert-annotation}

\textbf{Sampling YouTube Videos.} Since acquiring expert labels is expensive, we devised a stratified sampling method to select videos likely to contain myths. Unlike random sampling, which often yields irrelevant videos, our method aimed to create a targeted evaluation set with a balanced label distribution. Following prior works \cite{shen-etal-2023-modeling, park-etal-2024-valuescope}, we employed \texttt{GPT-4o} to predict labels\footnote{Note that perfect precision is not necessary since experts will subsequently annotate these videos.} (Table \ref{tab:annotation}) for videos in the \texttt{OUD Search Dataset}, then evenly sampled videos across labels to ensure that relevant videos to OUD are more likely to be chosen. See Figure \ref{appendix:sampling-prompt} for the prompt. %(e.g., ``Supports (1),'' ``Neutral (0),'' ``Opposes OUD Myths (-1)'')

\noindent\textbf{Creating the Gold Standard Dataset.} With the sampled videos, we conducted multiple rounds of annotations with six clinical researchers as experts using our annotation scheme (\S \ref{sec:refining-annotation}), resulting in a gold-standard dataset of 2,480 high-quality labels (8 myths$\times$310 videos). \edit{This represents a substantial annotation effort for a high-stakes health issue, exceeding or matching the scale of prior expert-annotated studies \cite{chandra-etal-2025-lived, oud-audit, info:doi/10.2196/30753}, and contributing a rich, high-quality source of annotations.} We detail annotation process in \S \ref{appendix:annotation-process}. Experts reported an average annotation time of 3 minutes per video.\footnote{\edit{The \texttt{OUD Search Dataset} averaged 7.17 minutes per video, but expert annotators often did not need to watch the full videos, especially when content was irrelevant to OUD.}}

Across six experts, we found Krippendorff's $\alpha$ score\footnote{We use Krippendorff's $\alpha$ as it allows for varying annotator counts, aligning with our setup.} of 0.76 for all annotations across myths on the 310 videos. The $\alpha$ score indicates a moderate agreement \cite{Krippendorff1980ContentAA} and is comparable to, or exceeds, the level of agreement reported in prior work \cite{ijcai2021p536, 10.1145/3582568}. See Table \ref{tab:krippendorff} for the complete list of $\alpha$ scores across myths. For each video, we used the label agreed upon by all experts. In cases of disagreements, the first author reviewed the expert annotations, watched the video, and assigned the final label. Table \ref{tab:label-distribution} provides the distribution of the expert-annotated labels in our gold standard dataset

\noindent\textbf{Consolidating from 4 to 3 Classes.} Given our focus on detecting OUD-related myths, we follow \citet{10.1145/3544548.3580846} and merge the ``neutral (0)'' and ``irrelevant (2)'' classes into a single ``neither (0)'' category, since they neither support nor oppose myths. This yielded a 3-class setup: supporting the myth (1), opposing the myth (-1), and neither (0).

%  scale the labeling process using automatic classifiers. 
\section{Labeling for Myths in YouTube Videos}\label{sec:scaling-labels}
With 164K videos in the \texttt{OUD Recommendation Dataset}, manual annotation by experts is infeasible. To scale the labeling of the OUD-related myths, we leverage LLMs for myth detection (\S \ref{sec:llm-myth-detection}), distill lightweight classifiers (\S \ref{sec:distillation-myth-detection}), and implement \methodname, an efficient triage pipeline in which the lightweight classifiers route challenging cases to the LLM (\S \ref{sec:inference-triage}). We apply \methodname to the \texttt{OUD Recommendation Dataset}, evaluating its efficiency and cost effectiveness (\S \ref{sec:applying-triage}).

\subsection{LLM-Based Myth Detection}\label{sec:llm-myth-detection}

For each myth, we used LLMs to label videos through the three-class classification task. The inputs consisted of text-based video metadata: title, description, transcript, and tags. Due to limited high-quality annotations (Table \ref{tab:label-distribution}) needed to fine-tune encoder-only models (e.g., \texttt{DeBERTA}), we used zero and few-shot prompting for in-context learning \cite{NEURIPS2020_1457c0d6}, which has shown strong performance in social science tasks against human experts  \cite{dammu-etal-2024-uncultured, tornberg2023chatgpt}.

We constructed a task-specific prompt for each myth. In the few-shot setting, we included five annotated examples from our gold-standard dataset,\footnote{We exclude few-shot examples from evaluation to avoid data leakage, resulting in 305 expert-labeled videos per myth.} following prior work \cite{oud-audit}. To compare performance, we evaluated 10 widely-used LLMs: two each from OpenAI, Anthropic, Google, and Meta, plus models from DeepSeek and Qwen (see Table \ref{tab:model-performance-myths-f1-score} for the models). We discuss the input features and prompt design considerations in \S \ref{appendix:llm-consideration} and show prompts in Figures \ref{fig:zero-shot-prompt}-\ref{fig:few-shot-prompt}.

\noindent\textbf{Results.} Tables~\ref{tab:llm-performance-part1}–\ref{tab:llm-performance-part3} present the full evaluation of 10 LLMs, with details in \S\ref{appendix:llm-evaluation}. \texttt{GPT-4o} with few-shot prompting consistently outperformed other models, achieving 0.82–0.87 macro F1-scores across myths (Table \ref{tab:myth-performance-summarized}). These results validate the effectiveness of LLMs for our task, matching or even exceeding prior works \cite{jung2025algorithmicbehaviorsregionsgeolocation, nguyen2024supportersskepticsllmbasedanalysis}. While \texttt{GPT-4o} offers a strong, scalable alternative to expert annotations and labels the remaining \texttt{OUD Search Dataset}, the API costs make it impractical for labeling the 164K-video \texttt{OUD Recommendation Dataset}, motivating the need for a lightweight model.

\begin{table}[t!]
\centering
\scriptsize
\setlength{\tabcolsep}{4.5pt}
\begin{tabular}{cccccc}
\toprule
\textbf{Myth} & \textbf{\texttt{GPT-4o}} & \textbf{\texttt{DeBERTA}} & \textbf{MSP} & \textbf{VET} & \textbf{MSP+VET} \\
\midrule
M1 & \textbf{0.87} (1) & 0.77 (0) & 0.81 (0.31) & 0.84 (0.53) & 0.86 (0.60) \\
M2 & \textbf{0.85} (1) & 0.70 (0) & 0.72 (0.10) & 0.79 (0.53) & 0.80 (0.57) \\
M3 & \textbf{0.86} (1) & 0.76 (0) & 0.82 (0.31) & 0.82 (0.52) & \textbf{0.86} (0.67) \\
M4 & \textbf{0.82} (1) & 0.62 (0) & 0.66 (0.04) & 0.76 (0.30) & 0.76 (0.31) \\
M5 & \textbf{0.82} (1) & 0.60 (0) & 0.63 (0.13) & 0.67 (0.23) & 0.68 (0.28) \\
M6 & \textbf{0.86} (1) & 0.76 (0) & 0.80 (0.20) & 0.80 (0.46) & 0.83 (0.52) \\
M7 & \textbf{0.85} (1) & 0.74 (0) & 0.80 (0.15) & 0.79 (0.37) & 0.81 (0.44) \\
M8 & \textbf{0.87} (1) & 0.78 (0) & 0.78 (0.00) & 0.81 (0.05) & 0.81 (0.05) \\
\bottomrule
\end{tabular}
\caption{Macro F1-scores across myths using the best-performing LLM (\textbf{\texttt{GPT-4o}}), the distilled model (\textbf{\texttt{DeBERTa-v3-base}}), and \methodname: maximum softmax probability (\textbf{MSP}), validation error tendencies (\textbf{VET}), and \textbf{MSP+VET}. Each row reports performance on 305 expert-annotated videos. Parentheses indicate the proportion of examples handled by \texttt{GPT-4o}---lower is better, reflecting greater reliance on the lightweight model and reduced reliance on larger, expensive models.}\vspace{-2.5mm}
\label{tab:myth-performance-summarized}
\end{table}

\subsection{Distillation for Myth Detection}\label{sec:distillation-myth-detection}

While \texttt{GPT-4o} performs well, its financial and computational costs make the model impractical for large-scale labeling of our task. Meanwhile, we lack sufficient high-quality, expert-labeled data to fine-tune a model (Table \ref{tab:label-distribution}). To address these challenges, \edit{we adopt a distillation approach, as prior works have shown that student models can be effectively trained from high-performing teacher models \cite{rao-etal-2023-makes, park-etal-2024-valuescope, 10.5555/3666122.3668142}. As discussed in \S \ref{sec:llm-myth-detection}, \texttt{GPT-4o} achieved the strongest performance on our task (macro F1: 0.82-0.87 across myths), making it a suitable teacher model. We therefore use \texttt{GPT-4o} to generate high-quality synthetic labels} \cite{10.5555/3666122.3668142} for the 1,466 videos in \texttt{OUD Search Dataset} that were not annotated by experts and train a lightweight student model on this synthetic data. 

This approach minimizes API and computational costs and avoids the instability of relying on proprietary, closed-source LLMs, whose behaviors can change over time \cite{openai_changelog}, while achieving strong performance on our gold-standard dataset. For each myth, we train a student model, \texttt{DeBERTa-v3-base} \cite{he2021DeBERTAdecodingenhancedbertdisentangled}, with training and experimental details in \S \ref{appendix:distillation}. 

\noindent\textbf{Results.} Table~\ref{tab:myth-performance-summarized} shows that \texttt{DeBERTa-v3-base}, trained on \texttt{GPT-4o}-generated synthetic labels, achieves macro F1-scores between 0.60 and 0.78 across myths, with scores $\ge 0.75$ on four myths. \edit{These results demonstrate strong performance despite the models' smaller size (186M parameters) relative to \texttt{GPT-4o} and its usage of synthetic data, validating both the quality of the student model and the \texttt{GPT-4o}-generated labels.} The results highlight the models' effectiveness on a high-stake, complex video classification task and their suitability for large-scale labeling. Full evaluation metrics are in Table~\ref{tab:myth-performance-best}, with additional details in \S\ref{appendix:distillation-evaluation-details}.

\subsection{\methodname Implementation}\label{sec:inference-triage}

With \texttt{GPT-4o} offering stronger performance and the distilled model enabling efficient large-scale labeling, \methodname can combine their strengths---using the lightweight model for routine cases and deferring harder ones to the strong, but costly LLM to optimize both cost and performance. To decide which examples to defer, we use two strategies: (1) \textbf{Maximum Softmax Probability (MSP)}, which uses the predicted class softmax probability as a simple, effective proxy for model confidence \cite{10.5555/3295222.3295387}, deferring examples below a chosen MSP threshold, and (2) \textbf{Validation Error Tendencies (VET)}, which defers predictions from classes with low validation performance (e.g., class-specific F1 < 0.8). We also evaluated combining MSP and VET, \edit{deferring an example to the high-performing \texttt{GPT-4o} if either the MSP or VET condition is met}. We discuss alternative deferral strategies, threshold selection method, and results motivating MSP and VET in \S\ref{appendix:triage-details}.

\noindent\textbf{Results.} Table~\ref{tab:myth-performance-summarized} summarizes the performance of \methodname using MSP, VET, and MSP+VET.  Compared to \texttt{DeBERTA}, MSP improved macro F1 by an average of 0.036 ($\pm$0.02) while deferring 0–31\% of examples to \texttt{GPT-4o}. VET achieved greater gains, improving macro F1 by 0.069 ($\pm$0.034) while deferring 5–53\% of examples, reflecting its aggressive deferral strategy based on class-level performance. The combined MSP+VET approach yielded the best results, increasing macro F1 by 0.085 ($\pm$0.03) with 5–67\% of examples deferred. Using MSP+VET, the triage achieved macro F1-scores between 0.68–0.86; notably, on M3, it matched \texttt{GPT-4o}'s performance while only deferring 67\% of examples. These results demonstrate that \methodname not only improves performance over the distilled model but also offers a scalable solution for annotating large datasets.

\subsection{Applying \methodname}\label{sec:applying-triage}

We applied \methodname using MSP+VET to label the 164K-video \texttt{OUD Recommendation Dataset} across 8 myths, totaling 1.3 million annotations. Of these, only 70,777 predictions (5.4\%) were deferred to \texttt{GPT-4o}, with \texttt{DeBERTA} handling the rest. Below, we compare the estimated time and costs of \methodname to experts and full \texttt{GPT-4o} labeling. Detailed calculations, including estimated environmental cost savings, are in \S\ref{appendix:cost-analysis}. In \S \ref{appendix:additional-evaluation}, we validated \methodname on 100 additional videos, observing comparable performance (0.77-1 macro F1) to that on the gold-standard dataset. %(see Table \ref{tab:myth-performance-recommendation}). 

Having an expert label 1.3 million annotations would require $\sim$8,209 hours and cost \$59.5K,\footnote{We use the U.S. federal minimum wage as a lower bound \cite{us-min-wage}.} while \texttt{GPT-4o} labeling would take $\sim$1,240 hours and cost \$21.8K in API usage. In contrast, \methodname---including \texttt{DeBERTA} training---reduced total time to $\sim$300 hours and cost to \$1,281.94. This represents a 98\% financial cost reduction and 96\% time savings compared to expert labeling, and a 94\% financial cost reduction and 76\% time savings compared to \texttt{GPT-4o} labeling. These results demonstrate that \methodname offers a highly scalable, practical solution for cost-efficient large-scale labeling in high-stakes domains.

\section{Assessing Overall Stance and Myth Bias}

\subsection{Determining Overall Stance}\label{sec:resolving-overall}

To assign each video a single overall stance label across the eight myths, we used the following heuristic: videos with only supporting or supporting+neither labels were marked as supporting; only opposing or opposing+neither as opposing; and only neither as neither. For videos with both supporting and opposing labels (63 in \texttt{OUD Search}, 193 in \texttt{OUD Recommendation}), we combined manual annotation and LLM-as-a-judge. Two authors annotated and arrived at a consensus on 63 videos; showing high agreement, an author labeled an additional 63; the remaining 130 were labeled using \texttt{GPT-4.1}, which achieved 0.79 macro F1 against human annotations (Table~\ref{tab:stance-performance}; full details in \S\ref{appendix:overall-stance}).

\subsection{Quantifying Myth Bias}\label{sec:myth-bias}

Using the overall stance labels, we adapt the misinformation bias score from \citet{10.1145/3392854} to quantify myth prevalence in YouTube search results: $\frac{(s - o)}{(s + n + o)}$, where $s$, $o$, and $n$ denote the frequency of \textit{supporting (1)}, \textit{opposing (-1)}, and \textit{neither (0)} videos, respectively. Thus, the bias score is a continuous value ranging from -1 (all videos oppose myths) to +1 (all videos support myths). Positive scores indicate a lean toward myths, negative scores indicate a lean toward opposing myths. Higher scores suggest a greater myth prevalence.

\section{Analysis}

Understanding the prevalence of myths on online platforms like YouTube is helpful for public health officials and platform developers to inform interventions and combat the opioid crisis \cite{mit_actionable, eysenbach2009infodemiology}. We use the predicted labels to analyze the prevalence of myths in 2.9K search results and 343K recommendations, uncovering actionable insights at scale and thereby demonstrating the utility of \methodname. We discuss additional analysis in \S~\ref{appendix:additional-analysis}.

\subsection{Prevalence of Myths in Search Results}\label{sec:prevalence-analysis-search}

\begin{table}[t!]
\centering
\resizebox{\columnwidth}{!}{%
\begin{tabular}{lccccccccc}
\toprule
\textbf{Label} & \textbf{M1} & \textbf{M2} & \textbf{M3} & \textbf{M4} & \textbf{M5} & \textbf{M6} & \textbf{M7} & \textbf{M8} & \textbf{Over.} \\
\midrule
Oppose   & 0.15 & 0.23 & 0.14 & 0.16 & 0.11 & 0.16 & 0.11 & 0.04 & 0.30 \\
Neither  & 0.77 & 0.69 & 0.78 & 0.81 & 0.85 & 0.76 & 0.82 & 0.91 & 0.51 \\
Support  & 0.08 & 0.09 & 0.09 & 0.03 & 0.05 & 0.08 & 0.07 & 0.05 & 0.20 \\
\bottomrule
\end{tabular}%
}
\caption{Distribution of labels for each OUD-related myths and overall (Over.), based on the 2.9K search results from the \texttt{OUD Search Dataset}.}\vspace{-1.mm}
\label{tab:myth-ratio}
\end{table}

\noindent\textbf{Overall, nearly 20\% of search results support OUD-related myths} (Table \ref{tab:myth-ratio}). Across individual myths, 3\%-9\% of search results support myths, reflecting a consistent presence of myth-supporting content. Meanwhile, 30\% of search results oppose myths, with 4\%-23\% of content per myth countering. While opposing content is present, it could be insufficient to meaningfully offset the persistence of myth-supporting videos. 
This raises concerns about the quality of information on YouTube, where users may encounter inadequately challenged myths that can misinform decision-making around opioid use and treatment. These insights can inform YouTube's content moderation strategies to reduce exposure to harmful myths and help public health officials design targeted health campaigns to proactively counter misinformation.

\begin{figure}[!t]
  \centering
\includegraphics[width=0.78\linewidth]{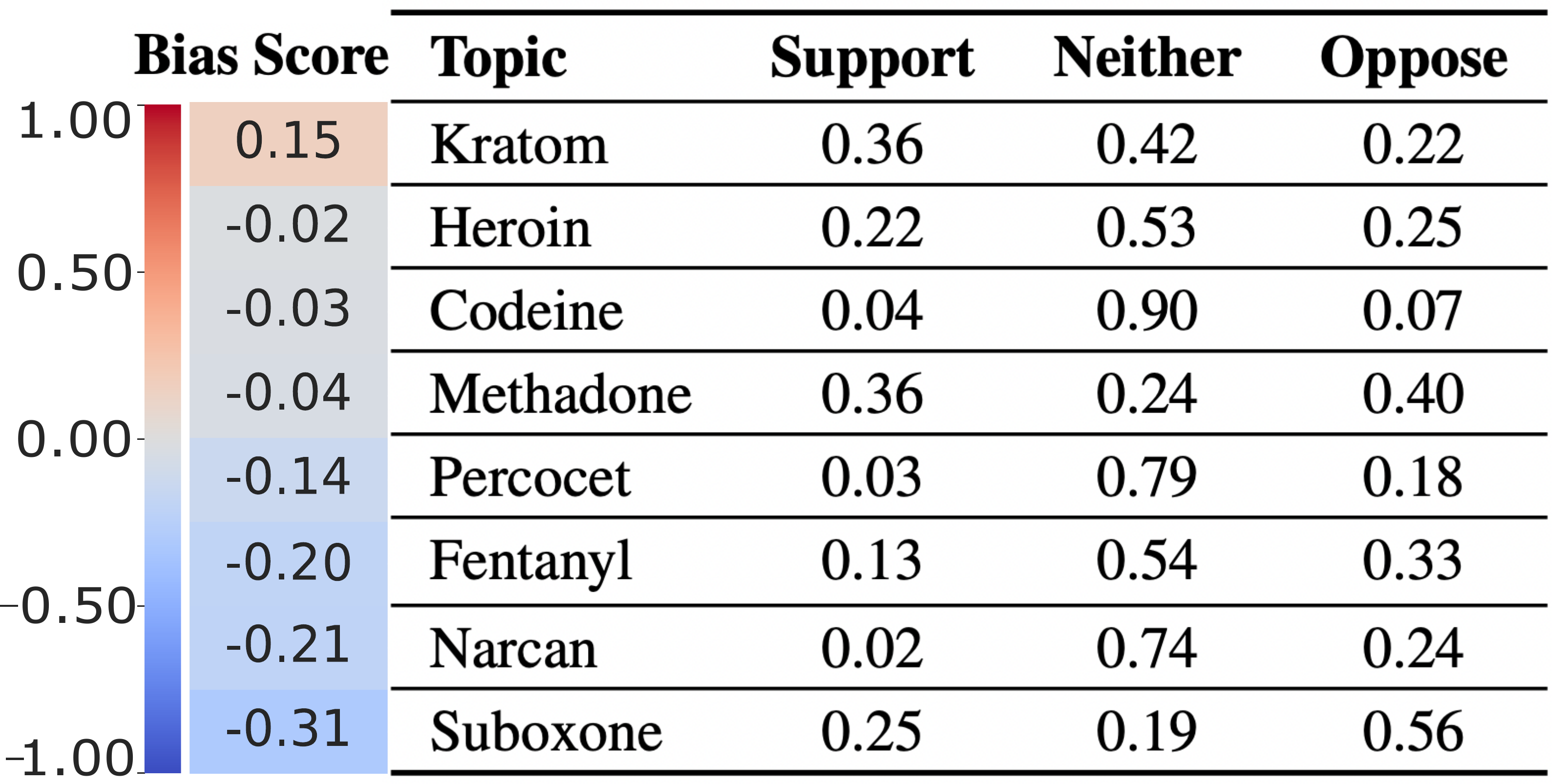}
  \caption{For each topic, we show the label distribution and myth bias score, computed using the overall stance labels from search results collected within the topic. Topics are sorted in descending order by bias score; higher values indicate greater prevalence of myths. }
  \label{fig:topic-wise-distribution}
\end{figure}

\noindent\textbf{Myth 2 shows the highest levels of support and opposition among all myths.} In Table \ref{tab:myth-ratio}, 9\% of search results support Myth 2, while 23\% oppose it. This myth is especially harmful as it frames OUD as a personal failure rather than a treatable medical disease, reinforcing stigma that people with OUD are weak or irresponsible. For example, one video states: \textit{``you're where you are because that's where you want to be''} (Table \ref{tab:oud-myths}). Such narratives can deter individuals from seeking treatment, reduce public support, and foster discrimination in social services \cite{tsai2019stigma, doi:10.1177/1178221816685087}. The high levels of supporting and opposing content suggest that this myth is contentious, motivating the need for targeted interventions. Platforms and officials can prioritize moderation and health campaigns to counter this myth; notably, LLM-based interventions show promise for increasing people's propensity toward MAT \cite{mittal2025exposurecontentwrittenlarge}.

\noindent\textbf{Kratom has the highest prevalence of myths across topics} (Figure \ref{fig:topic-wise-distribution}), with 36\% of search results supporting and only 22\% opposing OUD-related myths. This is concerning, given the widespread but debunked claims about Kratom's effectiveness in treating OUD. The high prevalence of myth-supporting content may mislead users towards unsafe alternatives \cite{mayo-clinic}, undermining evidence-based treatments like MAT. While Heroin and Methadone have slightly negative bias scores (-0.02 and -0.04), they still show high levels of myth-supporting content (22\% and 36\%). These findings can help platforms prioritize moderation on high-risk topics and inform public health officials where misinformation is most concentrated.

\begin{figure}[!t]
  \centering
\includegraphics[width=0.57\linewidth]{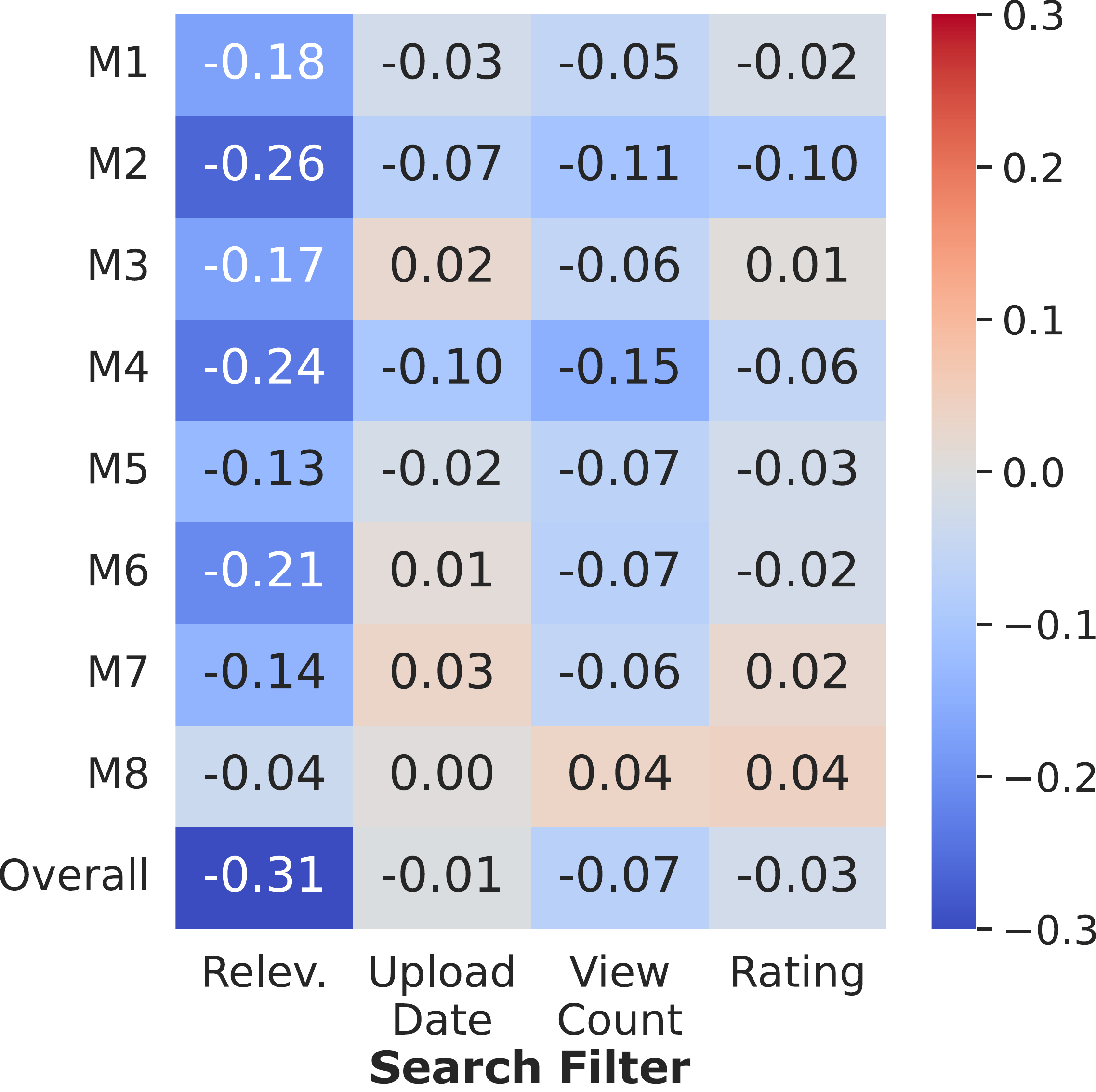}
  \caption{Bias scores for all 8 myths, including overall stance, and 4 search filters. Relevance (``Relev'') is YouTube's default sorting filter for search results.}
  \label{fig:filter-wise-distribution}
\end{figure}

\noindent\textbf{Switching from the default ``Relevance'' filter increases exposure to myths.} In Figure \ref{fig:filter-wise-distribution}, the ``Relevance'' filter consistently returns search results that lean towards opposing myths (bias scores from -0.04 to -0.26). In contrast, alternative filters---``Upload Date,'' View Count,'' and ``Rating''---consistently shift bias scores in a more positive direction, indicating increased prevalence of myth. This finding aligns with prior work in the domain of COVID-19 misinformation \cite{jung2025algorithmicbehaviorsregionsgeolocation}, and suggests that users seeking recent, most-viewed, or highly-rated videos (more likes than dislikes) are more likely to encounter myths. This is concerning, as users may place more trust in popular content, and those seeking the latest information about OUD may be at greater risk of encountering myths. These findings present an opportunity for YouTube to improve moderation by enhancing safeguards in non-default search filters.

\subsection{Prevalence of Myths in Recommendations}

\begin{figure}[!t]
  \centering
\includegraphics[width=0.93\linewidth]{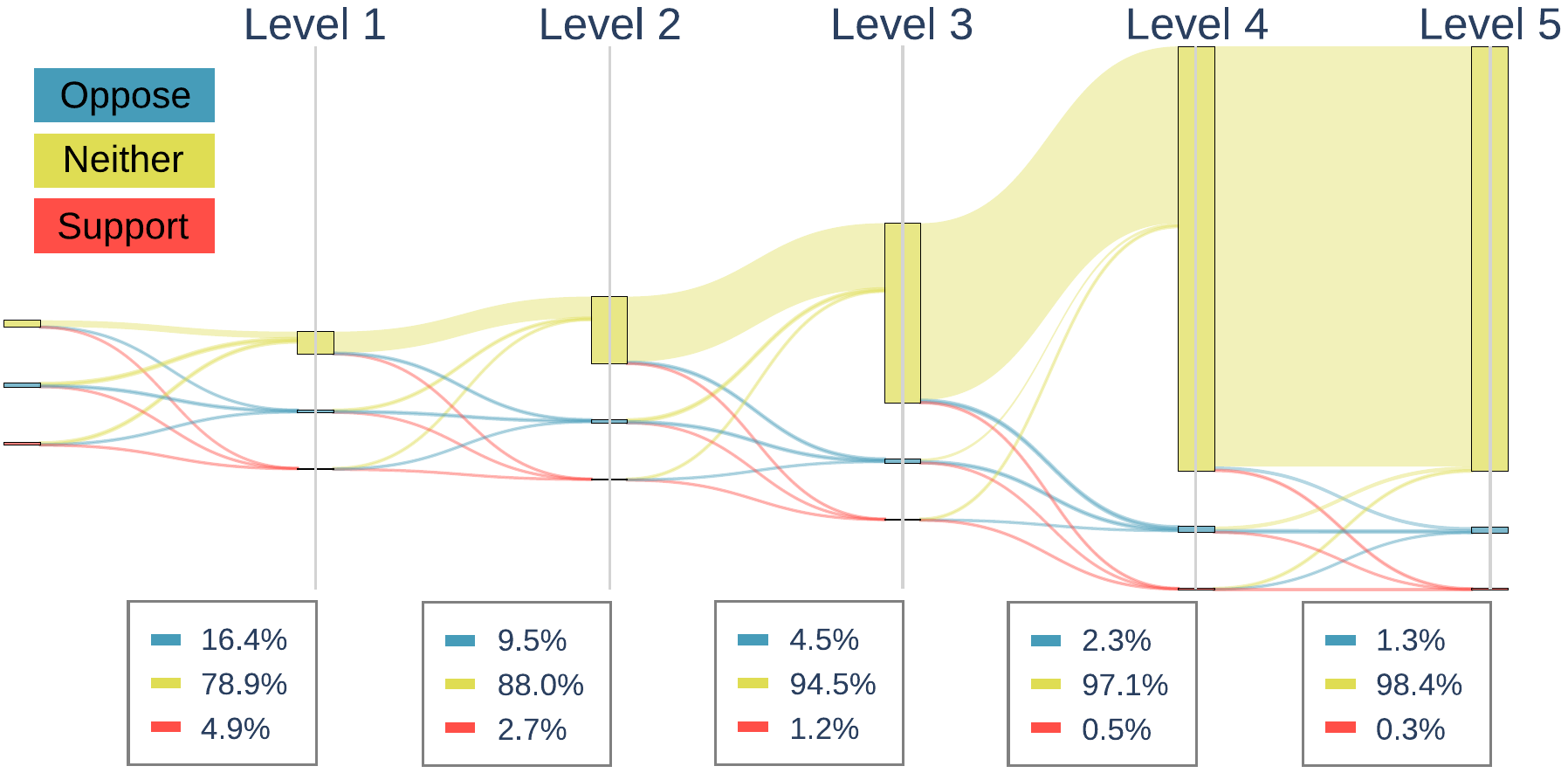}
  \caption{Recommendation transitions across levels. The edges between levels indicate transitions from a video's overall stance label to the labels of its recommended video. Node sizes increase across levels because more videos are recommended at each subsequent level. At the bottom, we display the distribution of overall stance labels within each recommendation level.}
  \label{fig:recommendation-transition}
\end{figure}

\noindent\textbf{Level 1 recommendations contained the highest proportion of myth-supporting videos (4.9\%)}, steadily declining to 0.3\% by level 5 recommendations. Similarly, myth-opposing content dropped from 16.4\% to 1.3\% across the same levels. As shown in Figure~\ref{fig:recommendation-transition}, the rise in ``neither'' labels suggests the recommendation algorithm increasingly surfaces unrelated content over time, aligning with prior findings that recommendations can play a moderating role when amplifying problematic content \cite{doi:10.1073/pnas.2313377121, Albadi_2022}. However, 4.9\% of level 1 recommendations supporting myths is concerning, as prior work shows recommendations can shape user engagement and viewing trajectories \cite{10.1145/3351095.3372879}. 

\noindent\textbf{At level 1, 12.7\% of recommendations to myth-supporting videos lead to other supporting videos}, rising to 22.2\% by level 5 (Appendix Table \ref{tab:recommendation-supporting-percent}). While supporting videos are not the most prevalent overall (Figure~\ref{fig:recommendation-transition}), this trend suggests that continued engagement with such videos in the recommendations increases exposure to more supporting videos. Additionally, 5.43\% of recommendations from opposing videos lead to supporting content at level 1, decreasing slightly to 3.25\% by Level 5, suggesting limited but persistent exposure even from opposing sources. These findings can inform YouTube's moderation efforts by helping identify recommendation pathways that may expose and reinforce users to myths.

\section{Conclusion \& Future Work}

We introduce \methodname, a scalable, cost-efficient pipeline for detecting 8 OUD-related myths across large video corpora. \methodname achieves strong performance on detecting OUD myths against expert labels, while greatly reducing annotation time and cost. Using \methodname, we present the first large-scale analysis of OUD myths on YouTube, revealing concerning levels of myth-supporting content and offering actionable insights for moderation and health interventions. 

By combining advances in NLP, public health, and clinical expertise, our work contributes a robust, extensible method for large-scale annotation in high-stakes domains like OUD, opening up many possibilities for applications and future research.

\noindent\textbf{Applications for Public Health.}\hspace{1mm}
\methodname can empower public health practitioners to monitor real-time misinformation trends, identify emerging myths, and launch targeted interventions promptly. Such insights can help clinicians understand common myths patients may encounter online, enabling better-informed, trust-building interactions.

\noindent\textbf{Platform Moderation and Auditing.}\hspace{-2mm}
\edit{Video-sharing platforms like YouTube have introduced policies to address harmful and unsubstantiated content \cite{google-policy2}, including medical misinformation policies targeting treatment-related claims \cite{google-policy}, indicating that misinformation mitigation is a platform priority.} \methodname can be integrated into platform moderation workflows to flag high-risk content, \edit{support scalable content auditing, and provide actionable insights that can inform at-scale moderation improvements.} Third-party researchers can use \methodname to evaluate how various algorithms and filter settings shape user exposure to misinformation at scale, informing platform transparency and algorithmic accountability.

\section{Limitations}
\noindent\textbf{Focus On Opioid Use Disorder.} We focus on OUD-related myths due to their high-stakes yet understudied nature on video-sharing platforms. However, there are several other important health domains, such as the COVID-19 pandemic \cite{jung2025algorithmicbehaviorsregionsgeolocation} and mental health \cite{nguyen2024supportersskepticsllmbasedanalysis}. Our methodology, such as \methodname and the experimental setup, can serve as blueprints for future work in other health domains

\noindent\textbf{Expanding Myths and Topics.} In this work, we examined 8 expert-validated OUD-related myths and 8 topics, consisting of 4 treatment and 4 opioid topics (\S \ref{sec:topics}). While this covers more ground than prior works \cite{oud-audit, info:doi/10.2196/30753}, many other myths, such as \textit{``It Is Expensive To Treat Patients With OUD''} \cite{additional-myth}, and topics (e.g., OxyContin) remain underexplored. Future works can extend \methodname and our annotation guidelines to cover a broader range of opioid-related myths and topics.

\noindent\textbf{Exploring Deeper and Beyond YouTube.} This work presents the first large-scale analysis of OUD-related myths on YouTube. While prior research highlights the role of personalization in amplifying problematic content \cite{10.1145/3392854}, future work can examine how personalization---such as user demographics, search history, and prior engagement--- shapes exposure to OUD-related myth. Other video-sharing platforms like BitChute \cite{nguyen2024supportersskepticsllmbasedanalysis} and TikTok \cite{10.1145/3485447.3512102} also merit investigation. Extending this analysis across platforms would enable cross-platform comparisons of myth prevalence. 

%this one reads more as a defense of our method rather than limitation. Why not end with things like crowd-sourced audit or privacy-preserving audits as next steps for future work.
\noindent\textbf{Google \textit{Trends} Alignment.} \edit{To curate search topics and queries, we use Google \textit{Trends}, configured to reflect YouTube Search, to capture highly popular opioid-related queries based on real-world search interests. This ensures our analysis reflects the amount of misinformation many YouTube users are likely to encounter when seeking OUD information, but may bias the dataset toward mainstream searches and underrepresent niche queries, including those used by stigmatized or vulnerable populations. Future work could address this gap through \textit{crowdsourced} audit \cite{10.1145/3544548.3580846} or privacy-preserving audits to capture how real users, including those from vulnerable populations, encounter and engage with OUD-related misinformation.}

\noindent\textbf{Scope of Language and Context.} While our method, including \methodname, is not limited to any specific language or contexts, our use of English queries on YouTube likely contributed to the lack of non-English content in our dataset (Table \ref{tab:label-distribution}). As the study centers on the U.S., where opioid overdose is a leading cause of death \cite{nihDrugOverdose}, it reflects a Western-centric context. Given the growing opioid crisis in other regions, such as Bolivia and Guyana \cite{health2023opioid}, future works can adapt \methodname to other languages and contexts to support more in-depth cross-cultural analyses of OUD-related myths online.

\noindent\textbf{More Extensive Data Collection.} In our work, we collected 2.9K search results and 343K recommendations, providing the first large-scale analyses on the OUD myth prevalence on YouTube. Future work can expand this by collecting data over longer periods to capture longitudinal trends and better understand how myths evolve over time.

\noindent\textbf{Model Misclassification.} We employ LLMs and distill lightweight classifiers validated on clinical expert-labeled datasets. Despite extensive model training and experimentation, the error rates in our \methodname pipeline may potentially influence \edit{the distillation of the lightweight classifiers and} our downstream analysis. While our text-based approach using models like \texttt{DeBERTA-v3} balances performance and efficiency (\S \ref{sec:inference-triage}), future work can improve performance by incorporating multimodal features (e.g., thumbnails, video frames) \cite{10.1145/3697349}, \edit{though this would increase computational and financial costs of the pipeline by requiring multimodal LLMs}, and integrating additional deferral mechanisms between experts and LLMs using uncertainty metrics from \citet{farr2024llmconfidenceevaluationmeasures}. \methodname can be adapted to use small, open-weight LLMs such as \texttt{Llama-3-8B} \cite{zhan-etal-2025-slm} for potentially better performance, though even models with $\ge$ 1B parameters would substantially increase training and inference costs. 

\section{Ethical Considerations}\label{sec:ethics}

We avoided recruiting real-world users in the data collection to prevent potential contamination of their YouTube search and recommendation \cite{10.1145/3392854}. Instead, we used non-personalized tools, such as YouTube Data and InnerTube APIs.

% expert annotation vs. crowdsourced
To minimize exposure to harmful content, we did not involve crowd workers and instead worked with clinical researchers in substance use. Following \citet{kirk2022handling}, we conducted regular check-ins and debriefs to safeguard all data handlers. All analyzed videos were publicly available at the time of collection, and we only used video metadata (titles, descriptions, transcripts), without accessing user-specific data. In line with best practices \cite{doi:10.1177/20563051211019004}, we will release only video IDs and labels to ensure reproducibility while protecting user privacy. Practitioners can use these IDs with the YouTube Data API to retrieve metadata and hydrate the released dataset.

% using publicly accessible LLMs & will publicly release all artifact
We used publicly accessible LLMs to detect OUD-related myths. While our method enables scalable detection of myths, it is not intended to replace expert judgment. We emphasize that \methodname should be used as a research tool to support public health and inform platform moderation---with appropriate expert oversight. 

\section*{Acknowledgements}

This paper was supported by the National Institute of  Health’s NIDA grant DA056725-01A1 and the National Science Foundation grant CNS-2344925. We thank the Cavazos Lab at the Washington University School of Medicine for providing clinical research expertise and annotations in this study.

\bibliography{anthology, custom}

\appendix

\section{Details on Curating Topics and Queries}\label{appendix:search-topic}
Here, we provide additional details on \textit{Trends} configuration and keyword pairwise comparison results. In Table \ref{tab:query-set}, we also include the final set of 73 search queries spanning the 8 opioid-related topics employed in our study.

\subsection{Google \textit{Trends} Configuration.}\label{appendix:trends-config}
We configured \textit{Trends} to focus on: (1) the United States (US), given the context of the opioid epidemic; (2) the period from January 1, 2021 to December 31, 2023 to capture recent search patterns and obtain popular queries; and (3) the YouTube Search, aligning with the platform of interest. 

\textit{Trends} allows users to search keywords either as \textit{terms} or \textit{topics}. Searching as a \textit{term} returns results that match the exact words in the query, while searching as a \textit{topic} includes results for all terms with similar meanings. When selecting the OUD-related topics (\S \ref{sec:selecting-topics}), we searched the keywords on \textit{Trends} as \textit{terms} rather than \textit{topics}, as not all keywords had corresponding \textit{topics}. WHen selecting the search queries (\S \ref{sec:search-queries}), we searched each keyword as both a \textit{term} and, when available, \textit{topic} to collect their top-10 related search queries. If a keyword (e.g., ``Narcan'') lacked a \textit{topic}, we used a synonymous topic (e.g., ``Naloxone'') to gather their top-10 related search queries. 

\begin{table}[t]
\centering
\small
\setlength{\tabcolsep}{3pt}
\begin{tabular}{p{2.5cm}p{2.5cm}c}
\toprule
\textbf{Keyword Topic} & \textbf{Category} & \textbf{Win-Rate} \\
\midrule
\textbf{Fentanyl} & Opioid & 1.000 \\
\textbf{Percocet} & Opioid & 0.963 \\
\textbf{Heroin} & Opioid & 0.926 \\
\textbf{Codeine} & Opioid & 0.889 \\
Opium & Opioid & 0.852 \\
\textbf{Kratom} & Treatment & 0.815 \\
Morphine & Opioid & 0.778 \\
Opiate & Opioid & 0.667 \\
Opioid & Opioid & 0.593 \\
China White & Opioid & 0.593 \\
\textbf{Narcan} & Treatment & 0.593 \\
Norco & Opioid & 0.593 \\
\textbf{Suboxone} & Treatment & 0.556 \\
Oxycodone & Opioid & 0.519 \\
Hydrocodone & Opioid & 0.444 \\
Tramadol & Opioid & 0.407 \\
\textbf{Methadone} & Treatment & 0.407 \\
Hydrochloride & Opioid & 0.333 \\
Opioids & Opioid & 0.333 \\
OxyContin & Opioid & 0.296 \\
Acetaminophen & Prescriptions & 0.259 \\
Naltrexone & Treatment & 0.222 \\
Vicodin & Opioid & 0.185 \\
Naloxone & Treatment & 0.148 \\
Ibogaine & Treatment & 0.111 \\
Vivitrol & Treatment & 0.037 \\
Imodium & Prescriptions & 0.037 \\
\bottomrule
\end{tabular}
\caption{Keyword popularity ranking based on pairwise comparisons using \textit{Trends}. Win-rates were calculated across all possible pairwise comparisons among 28 opioid-related keywords. The table is ordered by win-rate, with fentanyl (1.00) being the most searched term and treatment medications (e.g., Vivitrol, ibogaine) generally ranking lower than opioid substances. We selected 8 keywords as topics: the top four opioid-related and top four treatment-related terms (bolded).}
\label{tab:opioid-popularity}
\end{table}

\subsection{Keyword Pairwise Comparison Results.}

In \S\ref{sec:topics}, we used \textit{Trends} to perform pairwise comparisons of 28 keywords based on relative search popularity and rank them by popularity. Since \textit{Trends} provides comparative scores between two terms, we calculate win-rates---the proportion of comparisons a keyword wins---to rank all keywords. See Table \ref{tab:opioid-popularity} for the list of the 28 keywords and their corresponding win-rates. 

Given our focus on detecting OUD-related misinformation, we selected 8 keywords as topics---the top four opioid and top four treatment keywords---capturing the most popular topics across both opioids and medication-assisted treatment. These opioid and MAT-related keywords are both widely-searched in the context of opioid use and treatment, which are often associated with OUD-related myths on social media platforms\cite{info:doi/10.2196/44726}.

\begin{table*}[t]
\scriptsize
\centering
\begin{tabular}{p{2.3cm}p{9cm}}
\toprule
\textbf{Topic} & \textbf{Search Queries} \\
\midrule
Fentanyl & fentanyl, overdose fentanyl, fentanyl drug, what is fentanyl, fentanyl documentary, \newline fentanyl crisis, fentanyl addict, fentailo, fentynal \\
\cmidrule{2-2}
Percocet (Oxycodone) & percocet, oxycodone, oxycontin, oxy, oxycotton, oxycotin \\
\cmidrule{2-2}
Heroin & heroin addict, on heroin, heroin, heroin drug, heroin addiction, herion \\
\cmidrule{2-2}
Codeine & codeine, codeina, codine, codiene, codein, codien, codeine pills \\
\cmidrule{2-2}
Kratom & kratom withdrawal, kratom, what is kratom, kratom review, red kratom, kratom extract, \newline best kratom, kratom tea, kratom high, kratom effects, kratom benefits, kratom psychosis \\
\cmidrule{2-2}
Narcan & narcan, narcan training, narcan video, narcan use, narcan overdose, how to use narcan, \newline nasal narcan, naloxone, narcan rescue, narcan saves life, narcan reaction \\
\cmidrule{2-2}
Suboxone & suboxone, suboxone withdrawal, how to take suboxone, taking suboxone, suboxone clinic, how does suboxone work, suboxone detox, suboxone high, suboxone taper, suboxone strips, what is suboxone, suboxone film \\
\cmidrule{2-2}
Methadone & methadone, methadone clinic, methadone withdrawal, methadone detox, what is methadone, \newline house methadone, methadone addiction, methadone high, methadone clinic experience, methadone nursing \\
\bottomrule
\end{tabular}
\caption{The final set of 73 search queries spanning 8 opioid-related topics employed in our study. For each topic, we utilized 6-12 search queries associated with the topic. Note that the top four represent opioid-related topics, while the bottom four represent treatment-related topics (e.g., MAT).}
\label{tab:query-set}
\end{table*}

\begin{table}[t]
\scriptsize
\centering
\begin{tabular}{ccc}
\toprule
\textbf{Level \#} & \textbf{\# Recommendations} & \textbf{\# Unique Videos} \\
\midrule
1 & 6,356 & 3,107 \\
2 & 12,412 & 8,489 \\
3 & 33,916 & 21,849 \\
4 & 87,224 & 55,248 \\
5 & 202,799 & 126,585 \\
\midrule
All levels & 342,707 & 164,085\\
\bottomrule
\end{tabular}
\caption{Number of recommendations and unique videos collected per recommendation level.}
\label{tab:recommendation_levels}
\end{table}

\begin{table*}[ht]
\centering
\small
\resizebox{\textwidth}{!}{%
\begin{tabular}{p{5.5cm}p{5.7cm}p{5.7cm}}  
\toprule
\textbf{Myth} & \textbf{Example (Supports the Myth)} & \textbf{Example (Opposes the Myth)} \\
\midrule

M1: Agonist therapy or medication-assisted treatment (MAT) for OUD is merely replacing one drug with another. 
& \textit{``being on [suboxone] and you know... it is an opioid so I don't count that as clean time''}   
& \textit{``Buprenorphine is a semi-synthetic opioid and... was later adopted for treatment of opioid use disorder because... it was so helpful in treating addiction''}  \\

\addlinespace

M2: People with OUD are not suffering from a medical disease treatable with medication, but from a self-imposed condition maintained through the lack of moral fiber. 
& \textit{``you're where you are because that's where you want to be''}  
& \textit{``he had a disease just like my mother-in-law currently has cancer... I realized just how much of a grip opiates have on the user and the user’s brain''}  \\

\addlinespace

M3: The ultimate goal of treatment for OUD is abstinence from any opioid use. 
& \textit{``I don't want to be stuck on [methadone] forever... it's not really going to help you... it's like a Band-Aid''}  
& \textit{``Methadone is one of the most effective forms of treatment for opioid use disorder, cutting overdose risk in half and proving more successful in long-term recovery than abstinence-only approaches''}  \\

\addlinespace

M4: Only patients with certain characteristics are vulnerable to addiction. 
& \textit{``Why are Autistic people more prone to addiction?''}  
& \textit{``No one is immune to addiction, no matter what you look like, no matter where you're from... Addiction can impact you the same way as everybody else.''}   \\

\addlinespace

M5: Physical dependence or tolerance is the same as addiction. 
& \textit{``I don't think my brain differentiates whether I'm on suboxone, methadone, heroin, fentanyl, oxys, perks. I don't think it knows the difference.''}   
& \textit{``Tolerance is not addiction, dependence is not addiction... addiction is a whole other category''}  \\

\addlinespace

M6: Detoxification (cold-turkey) for OUD is effective. 
& \textit{``You gotta tough it out... going cold turkey and still doing manual labor jobs... so fight through it. It can be done.''}  
& \textit{``Untreated with medications, people will relapse at extremely high rates''}   \\

\addlinespace

M7: You should only take medication for a brief period of time. 
& \textit{``the longer you stay trapped in something, the harder it is going to become to step away from that medically assisted drug [methadone]''}  
& \textit{``Suboxone is to addiction what medicine is to heart disease or what insulin is to diabetes... Depending on the addict, it could be something you always take.''}   \\

\addlinespace

M8: Kratom is a non-addictive and safe alternative to opioids. 
& \textit{``kratom offers a real potential benefit and it is safe in its natural form for consumers to use''}  
& \textit{``FDA issued an urgent advisory, stating in part evidence shows kratom similar effects to narcotics like opioids and carries risks of abuse, addiction, and in some cases death.''}  \\

\bottomrule
\end{tabular}}
\caption{List of 8 OUD-related myths investigated in our study. We identified these myths from prior literature and clinical sources on OUD (refer to \S\ref{sec:identify-myth}). Each row presents a myth alongside video transcript examples from our dataset that either support or oppose it. Note that these video transcript examples were extracted by \texttt{GPT-4o} as part of the labeling process (\S \ref{sec:llm-myth-detection}).}
\label{tab:oud-myths}
\end{table*}

\begin{table*}[ht]
\centering
\small
\resizebox{\textwidth}{!}{%
\begin{tabular}{p{1cm}p{3.5cm}p{7cm}p{7cm}}  
\toprule
\textbf{Scale} & \textbf{Annotation Label} & \textbf{Description} & \textbf{Example Video (Title)}\\
\midrule
-1 & Opposes OUD Myths & The video opposes, debunks, or provides countervailing context to an OUD myth. For example, a video citing health authority sources to debunk the myth that MAT is simply replacing one drug with another. Annotators should specify which myth is being opposed. & \textit{Living Proof: Methadone and Buprenorphine for Opioid Addiction: Joseph} \\

\addlinespace
0 & Neutral OUD Information & The video broadly covers anything related to the opioid epidemic without clearly supporting or opposing OUD myths. This includes news reports on overdoses, songs about the opioid crisis, or personal experiences without a definitive stance. & \textit{Opioid overdose treatment NARCAN demo with Montgomery County Fire \& Rescue Service | FOX 5 DC}  \\

\addlinespace
1 & Supports OUD Myths & The video promotes or affirms an OUD myth, such as claiming MAT substitutes one addiction for another. Annotators should specify which myth is being supported. & \textit{How to get off opiates, fent, methadone, subs, or KRATOM - sleep through withdrawals} \\

\addlinespace
2 & Irrelevant & The video is unrelated to opioid use disorder or the broader opioid epidemic. For example, entertainment content or unrelated topics. & \textit{"They're The Arrogant Jerks That Saved Your Life" | House M.D.} \\

\addlinespace
3 & URL Not Accessible & The video could not be accessed at the time of annotation due to deletion, privacy settings, or geographic restrictions. & --- \\

\addlinespace
4 & Non-English Video & The video is in a language other than English and cannot be interpreted by annotators. & \textit{STOP Resepkan Codeine untuk Batuk!} \\
\bottomrule
\end{tabular}}
\caption{Annotation labels used for video classification. For each label, we provide the numeric scale value, label description, and an example video title from our dataset.}
\label{tab:annotation}
\end{table*}

\section{Obtaining Expert Annotations}\label{appendix:annotation-appendix}
In this section, we explain the procedure for refining the annotation scheme, discuss the annotation guidelines, and the process of obtaining expert annotations for our gold standard dataset. 

\subsection{Refining Annotation Scheme}\label{sec:refining-annotation}

Developing the qualitative coding scheme for labeling videos required multiple iterations. An author with prior experience in health misinformation research initially sampled 50 videos from the \texttt{OUD Search Dataset}. Through repeated analysis and annotations, the author proposed a two-step annotation process. First, each video was labeled using a 6-point scale:  ``Opposes OUD Myths (-1),'' ``Neutral OUD information (0),'' ``Supports OUD Myths (1),'' ``Irrelevant (2),'' ``URL not accessible (3),'' and ``Non-English Language (4) (see Table \ref{tab:annotation} for examples). Second, if a video was labeled as ``opposing (-1)'' or ``supporting (1)'' a myth, annotators were required to specify which \textit{myth(s)} were involved.

To refine our annotation task, six clinical researchers with expertise in substance use disorders independently annotated 20 videos, extensively discussed, and provided feedback, which we used to further refine our annotation guidelines. %(\S \ref{appendix:annotation-guideline}).

\subsection{Annotation Guidelines}\label{appendix:annotation-guideline}

In our annotation task, experts assigned an annotation label to each YouTube video, extracted relevant excerpts from the video metadata, and provided a brief ($\sim$10-word) justification. Following prior work \cite{jung2025algorithmicbehaviorsregionsgeolocation, 10.1145/3544548.3580846}, they reviewed metadata in priority order: (1) video title and description, then (2) video content/transcript to understand the overall premise of the video.

\begin{figure*}[!ht]
  \centering
  \includegraphics[width=0.8\linewidth]{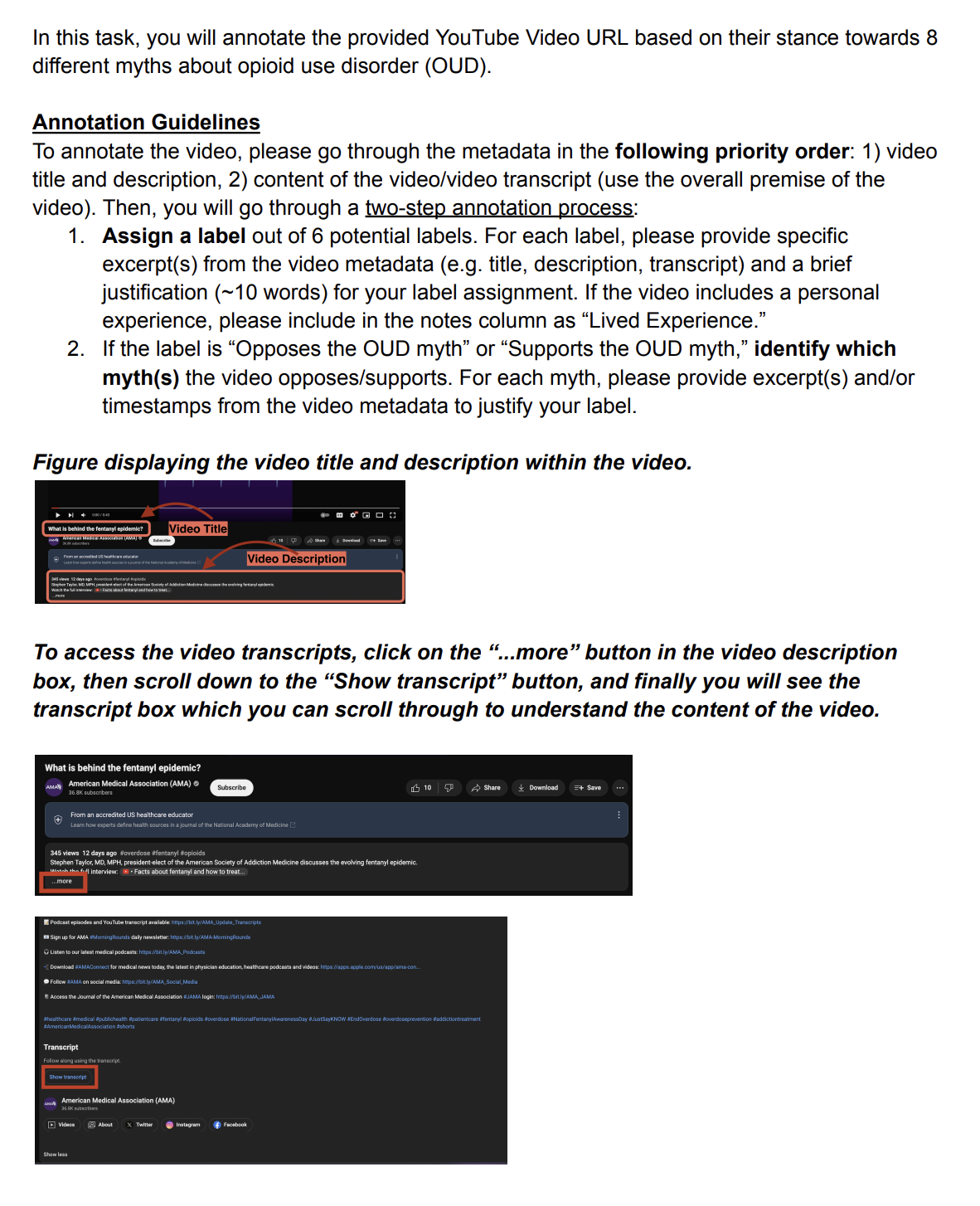}
  \caption{First page of the annotation guidelines provided to expert annotators.}
  \label{fig:annotation-guideline-page1}
\end{figure*}

\begin{figure*}[!ht]
  \centering
  \includegraphics[width=0.8\linewidth]{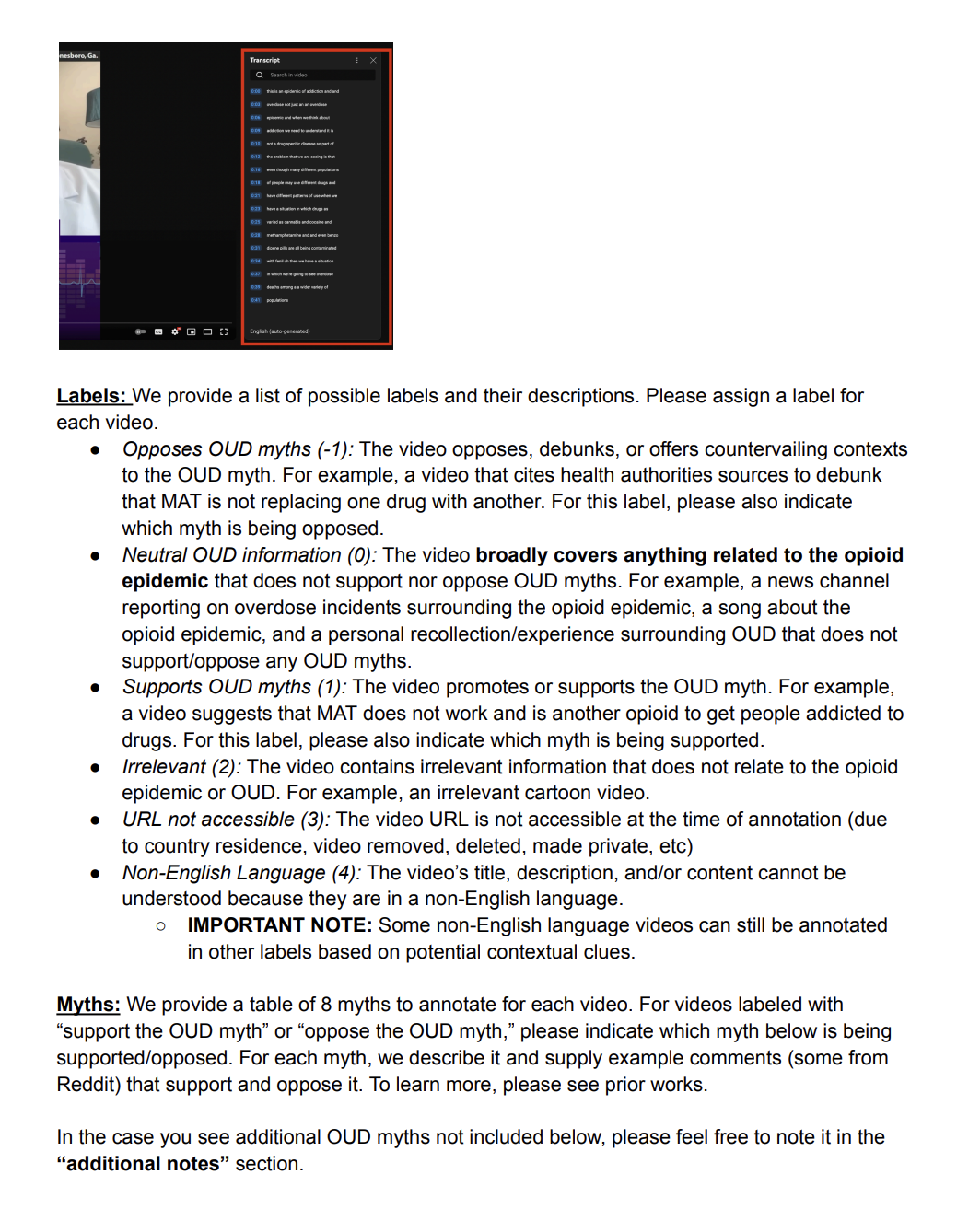}
  \caption{Second page of the annotation guidelines provided to expert annotators.}
  \label{fig:annotation-guidelien-page2}
\end{figure*}

After reviewing the video metadata, the experts followed a two-step annotation process. First, experts assigned one of six labels (Table \ref{tab:annotation}). Second, for videos labeled as ``Supports the OUD myths (1)'' or ``Opposes the OUD myths (-1),'' they identified the relevant myth(s) involved in the video, citing excerpts or timestamps to justify each myth (Table \ref{tab:oud-myths}). If a myth was not on the provided list, experts were instructed to note it explicitly. As shown in Figures \ref{fig:annotation-guideline-page1}-\ref{fig:annotation-guidelien-page2}, the annotation guidelines included the aforementioned instructions, labels, and myths. While refining and validating our annotation task, our task received positive feedback from the clinical researchers, who described it as a “straightforward coding task.” They found transcripts helpful and watched videos at 2x speed when transcripts were unavailable.

\begin{table}[!ht]
\centering
\renewcommand{\arraystretch}{1.2}
\begin{footnotesize}
\begin{tabular}{cc}
\toprule
\textbf{Myths} & \textbf{Krippendorff's $\alpha$} \\
\midrule
M1       & 0.777 \\
M2       & 0.689 \\
M3       & 0.728 \\
M4       & 0.687 \\
M5       & 0.770 \\
M6       & 0.670 \\
M7       & 0.687 \\
M8       & 0.806 \\
\midrule
\textbf{Overall} & \textbf{0.760} \\
\bottomrule
\end{tabular}
\end{footnotesize}
\caption{Krippendorff's $\alpha$ scores among six expert annotators across 310 video annotations on 8 myths and overall. The agreement scores are comparable to, or surpass, those reported in prior work.}
\label{tab:krippendorff}
\end{table}

\subsection{Annotation Process}\label{appendix:annotation-process}

To construct the gold-standard dataset, six clinical researchers, as experts, annotated 310 videos over three rounds of annotations. \edit{The clinical experts constituted a mixture of one clinical research scientist with a PhD, four graduate-level clinical researchers with Master’s degrees, and one undergraduate research assistant based at a flagship medical school in the U.S. These clinical researchers, including an undergraduate research assistant (with $>1$ year of research experience), regularly conduct clinical trials and perform research on substance use disorder and recovery, providing strong subject matter expertise.} We did not provide payment, but we obtained their consent to use their annotations for our study.

In the first round, all six experts independently annotated 20 videos, familiarizing themselves with the annotation guidelines (Appendix \ref{appendix:annotation-guideline}), providing feedback, and discussing disagreements. In the second round, they split into two groups of three annotators, annotating 40 videos per group. In the third round, they formed three pairs (e.g., groups of two), annotating 70 videos per group. Each round concluded with a debrief with the experts. Thus, in total, 210 videos had 2 annotators, 80 videos had 3 expert annotators, and 20 videos had 6 expert annotators. \edit{The annotation process spanned nearly three months and required close collaboration with clinical experts, who balanced these tasks alongside their own research.}

Among the six experts, we found Krippendorff's $\alpha$ score of 0.76 for all annotations across all 310 videos and 8 myths (see Table \ref{tab:krippendorff} for the complete list of scores per myth).  Despite the challenges of identifying OUD-related myths in text \cite{oud-audit}, our score indicates moderate agreement \cite{Krippendorff1980ContentAA}, and is comparable to the level of agreement reported in prior work \cite{ijcai2021p536, 10.1145/3582568, dammu-etal-2024-uncultured}. Table \ref{tab:label-distribution} contains the distribution of the expert-annotated labels across myths.

\begin{table*}[ht]
\centering
\small
\begin{tabular}{lcccccc}
\toprule
\textbf{Myth} & \textbf{Opposes (-1)} & \textbf{Neutral (0)} & \textbf{Supports (1)} & \textbf{Irrelevant (2)} & \textbf{URL Not Accessible (3)} & \textbf{Non-English (4)} \\
\midrule
M1 & 94 & 131 & 70 & 13 & 1 & 1 \\
M2 & 118 & 116 & 61 & 13 & 1 & 1 \\
M3 & 92 & 124 & 79 & 13 & 1 & 1 \\
M4 & 50 & 229 & 16 & 13 & 1 & 1 \\
M5 & 60 & 193 & 42 & 13 & 1 & 1 \\
M6 & 97 & 129 & 69 & 13 & 1 & 1 \\
M7 & 59 & 170 & 66 & 13 & 1 & 1 \\
M8 & 9  & 267 & 19 & 13 & 1 & 1 \\
\bottomrule
\end{tabular}
\caption{Expert-annotated label distribution per myth across the 310 annotated YouTube videos. Labels include: Opposes OUD Myths (-1), Neutral OUD Information (0), Supports OUD Myths (1), Irrelevant (2), Non-English Language (4), and URL Not Accessible (3). Given our focus on detecting OUD-related myths on YouTube, we follow prior works \cite{jung2025algorithmicbehaviorsregionsgeolocation, 10.1145/3544548.3580846} and merge the ``neutral (0)'' and ``irrelevant (2)'' classes into a single ``neither (0)'' category, since they neither support nor oppose OUD-related myths. This yields a 3-class classification task: supporting OUD-related myths (1), opposing OUD-related myths (-1), and neither (0). In our evaluation and analysis, we assign the ``URL not accessible'' class a score of 0, since we do not know their stance, thus providing a conservative estimate of OUD-related myths in our data.}
\label{tab:label-distribution}
\end{table*}

\begin{figure*}[!t]
\fbox{\begin{minipage}{0.95\textwidth}
\footnotesize
\ttfamily
\textbf{System Persona:} You are a public health expert with comprehensive knowledge of opioid use disorder (OUD) and the myths surrounding it, especially on social media platforms like YouTube.\\

You are tasked with carefully analyzing the provided video metadata to discern whether the provided YouTube video falls into one of four labels: opposes OUD myths, neutral OUD information, supporting OUD myths, and irrelevant.\\

Using the provided LABEL DESCRIPTIONS, please evaluate the YOUTUBE VIDEO METADATA and assign a label. Below, we provided the LABEL DESCRIPTIONS and define what kind of videos would fall into the category:\\

LABEL DESCRIPTIONS:
[TABLE-LABELS]\\

Potential OUD Myths:
[TABLE-MYTHS]\\

Note that these are not comprehensive and you may find other myths on opioid use disorders in the videos. Please include new potential myths in your justification.\\

Now, given what you learned from the LABEL DESCRIPTIONS above, please evaluate and assign a label to the YOUTUBE VIDEO METADATA and provide justification on your label with direct and concise EXCERPT(s) extracted from the YOUTUBE VIDEO METADATA. ONLY EXTRACT INTENTIONAL, BRIEF EXCERPTS TO JUSTIFY YOUR LABEL; DO NOT USE THE ENTIRE EXCERPT. FORMAT your response as a JSON object in the following structure [(LABEL, EXCERPTS, JUSTIFICATION)]. Make sure to have the keys LABEL, EXCERPTS, JUSTIFICATION in the JSON structure.\\

YOUTUBE VIDEO METADATA starts here *****:\\
Video Title: [TITLE]\\
Video Description: [DESCRIPTION]\\
Video Transcript: [TRANSCRIPT]\\
Video Tags: [TAGS]
\end{minipage}}
\caption{Zero-shot prompt used with \texttt{GPT-4o-2024-08-06} to assign preliminary labels to YouTube videos. These labels were then used for stratified sampling across labels, ensuring that relevant videos to OUD are more likely to be chosen for expert annotation. The prompt included the list of myths from Table~\ref{tab:oud-myths} and annotation labels and descriptions from Table~\ref{tab:annotation}.}
\label{appendix:sampling-prompt}
\end{figure*}

\section{Additional Details on LLM-Based Myth Detection}\label{appendix:llm-consideration}

Here, we discuss the feature descriptions, prompt design considerations, and evaluation results for LLM-based myth detection (\S \ref{sec:llm-myth-detection})

\subsection{Feature Descriptions.}\label{appendix:feature-description} In our prompts, we provide the following input features for the LLM. 

\begin{itemize}
[noitemsep,topsep=0pt,leftmargin=12pt]
    \item\textbf{Video Title: }The video's title.
    \item\textbf{Video Description: }A brief description of the video content.
    \item\textbf{Video Transcript: }Transcript contains the text of the video's content, from creator-provided or auto-generated subtitles, often reflecting the main themes discussed in the video.
    \item\textbf{Video Tags: }Descriptive keywords added by the content creators to help surface their video in search and recommendation \cite{youtube_tags}. 
\end{itemize}

\edit{Prior work \cite{jung2025algorithmicbehaviorsregionsgeolocation, 10.1145/3544548.3580846} has consistently shown that using all textual metadata--titles, descriptions, transcripts, and tags--yields the best performance for misinformation detection. For example, \citet{jung2025algorithmicbehaviorsregionsgeolocation} conducted ablation studies and found that removing any metadata component reduced performance in detecting COVID-19 misinformation in YouTube videos, indicating that each component contributes a useful signal. Based on these findings, we used the full set of text metadata components without re-running similar ablations, as our focus was on triage performance. }

\subsection{Prompt Design Considerations}

Our prompt design considerations were guided by OpenAI's prompt-engineering recommendations \cite{openai_promptengineering} and prior works \cite{githubtoxicity2022, dammu-etal-2024-uncultured, jung2025algorithmicbehaviorsregionsgeolocation, park-etal-2024-valuescope}. For each myth, we designed a zero-shot prompt (Figure \ref{fig:zero-shot-prompt}) and few-shot prompt (Figure \ref{fig:few-shot-prompt}) under these considerations. Below, we list the various prompt design features we considered:

\begin{itemize}
[noitemsep,topsep=0pt,leftmargin=12pt]
    \item\textbf{System Roles: } While personas can improve model performance \cite{openai_promptengineering2}, their effects are often unpredictable \cite{zheng-etal-2024-helpful}. However, \citet{zheng-etal-2024-helpful} suggests that ``gender-neutral, in-domain, and work-related roles'' can lead to better performance than other types of persona. Given our clinical and public health focus of our OUD myth detection task, we prompted the GPT models with the persona of a public health expert: \texttt{"You are a public health expert with specialized knowledge of opioid use disorder (OUD) and medication-assisted treatment (MAT)."} See Appendix Figures \ref{fig:zero-shot-prompt} and \ref{fig:few-shot-prompt} for the full persona. 
    \item\textbf{Contextual Details: }Since providing proper contextual details is helpful to LLMs to reason and justify their decisions \cite{openai_promptengineering}, we provide the definition of each myth (Table \ref{tab:oud-myths}) and descriptions of each label (Table \ref{tab:annotation}).
    \item\textbf{Temperature: }Temperature influences how models generate text \cite{openai_temperature}. Lower values (e.g., 0) makes the response more deterministic and consistent, while higher values produce more varied and creative outputs. Prior work \cite{githubtoxicity2022, dammu-etal-2024-uncultured, park-etal-2024-valuescope, jung2025algorithmicbehaviorsregionsgeolocation} found that a temperature of 0.2 performed best for deterministic tasks like misinformation and harmful language detection. Following this, we set the temperature to 0.2 for our task.
    \item\textbf{Zero-Shot vs. Few-Shot: }For each myth, we evaluated both zero-shot and few-shot prompting. Zero-shot prompts present the task without examples, while few-shot prompts provide examples to support in-context learning without updating model weights \cite{NEURIPS2020_1457c0d6}. For few-shot prompting, we manually crafted and provided five few-shot examples per myth, each containing a video title, description, transcript, and tags (see \S \ref{appendix:feature-description}), along with the assigned label (supporting, opposing, or neither) and their reasoning behind the provided label. 
    \item\textbf{Chain-of-Thought Reasoning: } Prompting LLMs to generate a chain of thought and justify their reasoning has been shown to improve performance in tasks \cite{10.5555/3600270.3602070}, including misinformation \cite{oud-audit, jung2025algorithmicbehaviorsregionsgeolocation} and harmful language detection \cite{dammu-etal-2024-uncultured}. Following this approach, we prompt the LLMs to output a label, extract a brief excerpt from the input video metadata, and provide a justification. To support full chain-of-thought reasoning, we set the output limit to 1024 tokens, allowing the model to generate without short output constraints.
\end{itemize}

\subsection{LLM Evaluation Results}\label{appendix:llm-evaluation}

Using both zero-shot and few-shot prompts, we evaluated 10 LLMs on the gold-standard dataset across 8 OUD-related myths. Detailed performance results of all 10 LLMs are shown in Tables \ref{tab:llm-performance-part1}, \ref{tab:llm-performance-part2}, and \ref{tab:llm-performance-part3}, with a summary of the best macro F1-scores by model in Table \ref{tab:model-performance-myths-f1-score}.

\texttt{GPT-4o-2024-08-06} consistently surpassed other models, particularly with few-shot prompts. It achieved macro F1-scores between 0.818–0.871 and accuracies between 0.849–0.977 across myths, validating the quality of our prompts and the effectiveness of using LLMs for our task. Its strongest result was for M1 (e.g., \textit{Agonist therapy or MAT for OUD is merely replacing one drug with another.}) with a macro F1-score of 0.871, and the weakest for M4 (e.g., \textit{Only patients with certain characteristics are vulnerable to addiction.}) at 0.818.

\texttt{Claude-3.5-Sonnet-20241022} performed comparably to \texttt{GPT-4o}, with macro F1-scores ranging from 0.741–0.864 and accuracies from 0.813–0.964. Despite their smaller sizes, \texttt{GPT-4o-mini-2024-07-18} and \texttt{Claude-3.5-Haiku-20241022} also demonstrated strong performance. Notably,  \texttt{GPT-4o-mini} performed well on M1, M3, and M6, and \texttt{Claude-3.5-Haiku} on M1, M3, M6, M7, and M8 (all $\ge$ 0.75 macro F1-scores), suggesting they may serve as cost-effective alternatives to their larger, more expensive counterparts for specific myths. 

\texttt{Meta-Llama-3-8B-Instruct}, the smallest model in our evaluation, had the weakest performance, with macro F1-scores ranging from 0.257–0.548 and accuracies from 0.382–0.721. This suggests that small language models may struggle to effectively detect misinformation without extensive fine-tuning \cite{zhan-etal-2025-slm}. Open-weight models like \texttt{DeepSeek-V3} and \texttt{Qwen-2.5-72B-Instruct} performed strongly, achieving macro F1-scores $\ge$ 0.75 on multiple myths---M1, M3, M6, M7, and M8 for \texttt{DeepSeek}; and M1, M2, M3, M6, M7, and M8 for \texttt{Qwen}. These strong performance suggests that open-weight models can offer competitive alternatives to proprietary LLMs for misinformation detection, especially in settings where transparency or customization is critical.

Few-shot prompting outperformed zero-shot prompting in nearly all cases. For example, \texttt{GPT-4o} saw macro F1-score improvements of 0.044–0.25 when using few-shot prompts compared to zero-shot prompts.

\begin{table*}[ht]
\centering
\small
\resizebox{\textwidth}{!}{%
\begin{tabular}{ccccccccccc}
\toprule
Myth & \texttt{GPT-4o} & \texttt{GPT-4o-mini} & \texttt{Claude-Sonnet} & \texttt{Claude-Haiku} & \texttt{Llama-3-8B} & \texttt{Llama-3.3-70B} & \texttt{DeepSeek} & \texttt{Gemini-Pro} & \texttt{Gemini-Flash} & \texttt{Qwen-72B} \\
\midrule
M1 & \textbf{0.871} & 0.808 & 0.864 & 0.860 & 0.509 & 0.765 & 0.845 & 0.824 & 0.728 & 0.829 \\
M2 & \textbf{0.854} & 0.690 & 0.818 & 0.717 & 0.333 & 0.759 & 0.728 & 0.692 & 0.679 & 0.791 \\
M3 & \textbf{0.859} & 0.752 & 0.839 & 0.804 & 0.548 & 0.747 & 0.809 & 0.807 & 0.665 & 0.790 \\
M4 & \textbf{0.818} & 0.628 & 0.741 & 0.578 & 0.340 & 0.630 & 0.587 & 0.561 & 0.588 & 0.617 \\
M5 & \textbf{0.824} & 0.667 & 0.743 & 0.675 & 0.318 & 0.707 & 0.734 & 0.707 & 0.683 & 0.716 \\
M6 & \textbf{0.857} & 0.791 & 0.832 & 0.807 & 0.376 & 0.767 & 0.838 & 0.830 & 0.791 & 0.764 \\
M7 & \textbf{0.853} & 0.747 & 0.792 & 0.797 & 0.504 & 0.772 & 0.766 & 0.751 & 0.637 & 0.810 \\
M8 & \textbf{0.866} & 0.680 & 0.758 & 0.860 & 0.409 & 0.766 & 0.809 & 0.752 & 0.754 & 0.792 \\
\bottomrule
\end{tabular}
}
\caption{The best model performances (Macro F1-Score) across 8 OUD-related myths. For each myth, we bolded the best model performance. For each myth, we evaluate the performance on 305 videos from the expert-annotated gold standard dataset, excluding the five few-shot examples used in the prompt. Note that \texttt{GPT-4o}: \texttt{GPT-4o-2024-08-06}, \texttt{GPT-4o-mini}: \texttt{GPT-4o-mini-2024-07-18}, \texttt{Claude-Sonnet}: \texttt{Claude-3.5-Sonnet-20241022}, \texttt{Claude-Haiku}: \texttt{Claude-3.5-Haiku-20241022}, \texttt{Llama-3-8B}: \texttt{Meta-Llama-3-8B-Instruct}, \texttt{Llama-3.3-70B}: \texttt{Meta-Llama-3.3-70B-Instruct}, \texttt{DeepSeek}: \texttt{DeepSeek-v3}, \texttt{Gemini-Pro}: \texttt{Gemini-1.5-Pro}, \texttt{Gemini-Flash}: \texttt{Gemini-1.5-Flash}, \texttt{Qwen-72B}: \texttt{Qwen-2.5-72b-instruct}. Please refer to Tables \ref{tab:llm-performance-part1}-\ref{tab:llm-performance-part3} for the full performance metrics using both zero-shot and few-shot prompts.}
\label{tab:model-performance-myths-f1-score}
\end{table*}

\section{Distillation for Myth Detection}\label{appendix:distillation}

\subsection{Training Details}\label{appendix:distillation-details}
For our distillation, we used  \texttt{DeBERTa-v3-base} model as the base model for our experiment. Prior works \cite{jung2025algorithmicbehaviorsregionsgeolocation, park-etal-2024-valuescope} employed \texttt{DeBERTa-v3-base} for misinformation detection task and model distillation, respectively, reporting strong performances in both tasks. We fine-tuned a separate model for each myth, resulting in 8 final models for our task. 

As detailed in \S \ref{sec:search-result-collection}, we collected 1,776 unique videos in our \texttt{OUD Search Dataset}, of which 310 were annotated by experts to form our gold-standard dataset. The remaining 1,466 videos were labeled by \texttt{GPT-4o-2024-08-06}, the best-performing LLM for our task as described in the previous section, through the 3-class classification task for each myth. These synthetic labels were split 80:20 into training\footnote{As discussed in \S\ref{sec:llm-myth-detection}, we excluded few-shot examples from evaluation to avoid data leakage and included these examples during training.} and validation sets, with the expert-annotated labels serving as the test set. Input features included concatenated video title, description, transcript, and tags (\S\ref{appendix:feature-description}), truncated to the first 1,024 tokens to fit model constraints—an approach shown to retain high performance in prior work \cite{jung2025algorithmicbehaviorsregionsgeolocation}.

We trained models using the Adam optimizer and cross-entropy loss. We conducted a grid search over learning rates (5e-6, 1e-5, 1e-6) and weight decays (5e-4, 1e-4, 5e-5), with batch size (8) and epochs (20) fixed. Some myths in the training data exhibited class imbalance, which can hinder model performance across underrepresented classes. To address this, we tested data balancing strategies such as upsampling and class-weighted loss, which have shown effectiveness in prior work \cite{Buda_2018}. In \S\ref{appendix:llm-augmentation}, we also experimented with LLM-based data augmentation to expand and balance the training data, given its potential to outperform traditional augmentation techniques. However, due to only marginal gains and high API costs, we prioritized simpler upsampling and class-weighted loss methods for training across the myths.

For each myth, we selected hyperparameters based on macro F1-score on the validation set and chose the final model based on the test macro F1-score. All models were trained on a single NVIDIA A40 GPU with 48GB of memory, and each training session (20 epochs) took approximately 60 minutes.

\subsection{Experimenting with LLM Data Augmentation}\label{appendix:llm-augmentation}

Prior work has shown that LLMs can effectively augment and generate synthetic data, often outperforming traditional augmentation methods like back-translation and lexical substitution \cite{jahan2024comprehensivestudynlpdata, nakada2025syntheticoversamplingtheorypractical}. To address class imbalance in our dataset and expand our training data, we use LLMs to generate synthetic examples grounded in existing training data.

\subsubsection{Experiment Setup}

We focus on Myth 4 (e.g., Only patients with certain characteristics are vulnerable to addiction), which had poor performance and a severe class imbalance: 25 supporting, 1,261 neither, and 180 opposing examples among the 1,466 videos labeled by \texttt{GPT-4o-2024-08-06} in the previous section. To balance the classes and expand the training data, we use \texttt{GPT-4.1-2025-04-14} to generate synthetic metadata for the “supporting” and “opposing” classes, conditioning the generation on example metadata from the training set to match tone, structure, and content (see Figure~\ref{fig:llm-augmentation-prompt}). 

We follow prior work \cite{oud-audit} and set the temperature to 0.7. In total, we generate 500 supporting examples (20 synthetic examples per 25 original) and 360 opposing examples (2 synthetic examples per 2 original). 
%We qualitatively reviewed the outputs to validate the quality of the generated data. For example, consider the following generated video title from a supporting example: \textit{Are Only Certain People at Risk for Opioid Addiction? The Truth Revealed.}

%\begin{itemize}
%[noitemsep,topsep=0pt,leftmargin=12pt]
%    \item\textbf{Video Title: }\textit{Are Only Certain People at Risk for Opioid Addiction? The Truth Revealed}
%    \item\textbf{Video Description: }\textit{Some people just aren\'t built for addiction—others are. We dive into what makes certain individuals more vulnerable to opioid use disorder, and why not everyone is at risk. Discover the traits, backgrounds, and circumstances that set apart those who become addicted from those who don\'t. »»» Subscribe for more real stories and expert insight on opioid use disorder.}
%    \item\textbf{Video Transcript: } \textit{When it comes to opioid addiction, not everyone is at risk in fact, research and real-life stories show that only people with certain characteristics are truly vulnerable to developing an addiction I sat down with Dr. James Porter, an addiction specialist, who explained the risk factors in detail some people have what's called an "addictive personality"—they might be more impulsive or have a family history of substance abuse...}
%    \item\textbf{Tags: } \textit{"opioid addiction", "addictive personality", "risk factors", "opioid use disorder", "OUD", "vulnerability to addiction", "genetics and addiction", "mental health", "addiction stories", "opioid crisis"}
%\end{itemize}

\begin{table}[t!]
\centering
\small
\begin{tabular}{lcc}
\toprule
\textbf{Setting} & \textbf{Accuracy} & \textbf{Macro F1} \\
\midrule
Base & 0.790 & \textbf{0.622} \\
Class Weight Loss & 0.790 & 0.603 \\
Upsample & 0.748 & 0.546 \\
LLM Data Augmentation & \textbf{0.816} & 0.613 \\
\bottomrule
\end{tabular}
\caption{Performance of  \texttt{DeBERTa-v3-base} on the three-class classification task for Myth 4, evaluated on the test set based on expert-annotated labels under the base supervised setup and various data balancing strategies (e.g., upsampling, class-weighted loss, and LLM-based data augmentation).}
\label{tab:m4-balancing}
\end{table}

\subsubsection{Experimental Results}

Using the LLM augmented data, we trained  \texttt{DeBERTa-v3-base} following the training procedure in \S \ref{appendix:distillation-details}. As baselines, we compared models trained with upsampling, class-weighted loss function, and a base supervised setup without these techniques. 

Table \ref{tab:m4-balancing} reports the performance on the three-class classification task for Myth 4, evaluated on the test set using expert-annotated labels. The base setup achieved the highest macro F1-score (0.622), while LLM-based augmentation yielded the highest accuracy (0.816). Given the class imbalance, macro F1 is a more informative metric, as it reflects balanced performance across all classes. These results are consistent with prior works \cite{choi2025limited, kruschwitz-schmidhuber-2024-llm}, which found that LLM-generated data often yields minor improvements in misinformation and toxicity detection. Given the minor performance improvements and the API costs of generating synthetic data, we focused on the base, class-weighted loss, and upsampling strategies when training models across the remaining myths.

\begin{figure*}[!t]
\centering
\fbox{\begin{minipage}{0.95\textwidth}
\footnotesize
\ttfamily
\textbf{System Persona:} You are a content creator on YouTube on opioid use disorder (OUD) and medication-assisted treatment (MAT).\\

You are given a myth and an example of video metadata. Your task is to generate new video metadata that [STANCE] the provided MYTH. Match the tone and style of the example metadata as closely as possible. Keep the total output under 1000 words.\\

Format your output as a JSON object, where each key is a video metadata field and the corresponding value contains the generated metadata. Each metadata entry must include the fields: TITLE, DESCRIPTION, TRANSCRIPT, and TAGS.\\

***MYTH TO REFERENCE STARTS HERE. Note that the generated metadata should [STANCE] the myth.***\\
MYTH: [MYTH]\\
***MYTH TO REFERENCE ENDS HERE***\\

***REFERENCE VIDEO METADATA STARTS HERE***\\
TITLE: [TITLE]\\
DESCRIPTION: [DESCRIPTION]\\
TRANSCRIPT: [TRANSCRIPT]\\
TAGS: [TAGS]\\
***REFERENCE VIDEO METADATA ENDS***
\end{minipage}}
\caption{Prompt used to generate synthetic video metadata for a given myth and stance (e.g., ``suppport''  or ``oppose''), conditioned on example metadata to guide tone and structure.}
\label{fig:llm-augmentation-prompt}
\end{figure*}

\begin{table}[t!]
\centering
\small
\setlength{\tabcolsep}{2pt}
\begin{tabular}{cccccc}
\toprule
\textbf{Myth} & \textbf{Train Acc.} & \textbf{Val Acc.} & \textbf{Val F1} & \textbf{Test Acc.} & \textbf{Test F1} \\
\midrule
M1  & 0.98 & 0.93 & 0.74 & 0.79 & 0.77 \\
M2  & 0.96 & 0.88 & 0.72 & 0.73 & 0.70 \\
M3  & 0.98 & 0.90 & 0.74 & 0.78 & 0.76 \\
M4  & 0.97 & 0.90 & 0.61 & 0.79 & 0.62 \\
M5  & 0.99 & 0.94 & 0.71 & 0.76 & 0.60 \\
M6  & 0.99 & 0.94 & 0.81 & 0.79 & 0.76 \\
M7  & 1.00 & 0.93 & 0.74 & 0.80 & 0.74 \\
M8  & 0.97 & 0.95 & 0.70 & 0.97 & 0.78 \\
\bottomrule
\end{tabular}
\caption{The best performance results achieved by \texttt{DeBERTa-v3-base} on the three-class classification task across myths. The training accuracy (Train Acc.), validation accuracy (Val Acc.), and validation macro F1-score (Val F1) are based on the \texttt{GPT-4o}-generated synthetic labels. The test accuracy (Test Acc.) and test macro F1-score (Test F1) are based on expert-annotated labels. All test results are evaluated on the same 305 videos from the expert-annotated gold-standard dataset per myth.}
\label{tab:myth-performance-best}
\end{table}

\subsection{Evaluation Results}\label{appendix:distillation-evaluation-details}

Table \ref{tab:myth-performance-best} shows the evaluation results for the trained \texttt{DeBERTa-v3-base} models across the 8 myths. The validation accuracy and macro F1-scores (Val) are computed using a held-out set with \texttt{GPT-4o} generated labels, and the test accuracy and macro F1-scores (Test) are computed using the expert annotations from the gold-standard dataset (\S \ref{sec:expert-annotation}).\footnote{As in \S\ref{appendix:llm-consideration}, test performance is evaluated on the same 305 expert-annotated videos from the gold standard dataset per myth, excluding the five few-shot examples used in LLM prompts.} The models perform reasonably well, achieving test macro F1-scores between 0.60 and 0.78 across myths. The close alignment between validation and test F1-scores further supports the high quality of the \texttt{GPT-4o}-generated labels.

\section{Additional Details on \methodname}\label{appendix:triage-details}

To efficiently scale high-quality video classification while managing cost, we propose \methodname: a lightweight model handles confident predictions, and uncertain ones are deferred to \texttt{GPT-4o}. We evaluate four deferral strategies to estimate model uncertainty:

%Prior work has explored \textit{confidence-based cascades}, where models handle “easy” examples quickly and defer uncertain cases to a stronger, more complex model when the current model's confidence is low \cite{10.5555/3666122.3666553}. 

\subsection{\methodname Considerations}\label{appendix:triage-considerations}
For \methodname, we considered four deferral approaches: (1) Maximum Softmax Probability (MSP), which defers predictions with low confidence based on the predicted class' softmax probability \cite{hendrycks2017a}; (2) Validation Error Tendencies (VET), which defers predictions from classes with low validation performance, indicating systematic model weakness; (3) Monte Carlo Dropout (MC-Dropout), which estimates uncertainty via 20+ forward passes per input to capture prediction variability from dropped neurons in the model \cite{gal2016dropout}; and (4) Softmax Entropy, which defers predictions with high entropy in the softmax probability distribution (indicating greater uncertainty). Prior works have shown that MSP is a strong method for selective predictions and model cascading \cite{varshney-baral-2022-model, geifman2017selectiveclassificationdeepneural}

\subsection{Methodological Details}\label{appendix:triage-methods}
For each method, we determine the optimal deferral threshold using the validation set, excluding examples that do not meet the threshold, computing the macro F1-score, and selecting the threshold value that maximizes macro F1-score based on the \texttt{DeBERTA-v3} predictions.

\begin{itemize}
[noitemsep,topsep=0pt,leftmargin=12pt]
    \item\textbf{MSP.} We use grid search to find the optimal softmax probability threshold for deferring predictions to \texttt{GPT-4o}. Thresholds from 0 to 1 (in 0.01 increments) are evaluated on the validation set. At inference time, any predictions with a softmax probability value that fall below the threshold are deferred to \texttt{GPT-4o}.
    \item\textbf{VET.} We compute per-class F1-scores on the validation set and identify low-performing classes with a per-class F1-score < 0.8. Prior works have often found F1-scores $\ge 0.8$ for detecting misinformation in text \cite{jung2025algorithmicbehaviorsregionsgeolocation, oud-audit}, motivating our use of this value to trigger deference in the VET strategy.\footnote{Future works can adjust this threshold, lowering it for a more relaxed deferral or increasing it for a more aggressive deferral strategy.} At inference time, any prediction falling into these low-performing classes is deferred to \texttt{GPT-4o}. 
    \item\textbf{MC-Dropout.} For each example, we conduct 20 forward passes with an active dropout layer. This produces 20 mean class probability distribution per input, capturing model uncertainty through output variability. We compute the entropy of the predicted class probabilities and use it as an uncertainty score. We perform a grid search to find the optimal entropy thresholds for deferring predictions to \texttt{GPT-4o}. Thresholds from 0 to 1 (in 0.01 increments) are evaluated on the validation set. At inference time, any predictions with an entropy value that is above the threshold are deferred to \texttt{GPT-4o}.
    \item\textbf{Softmax Entropy.} Like MSP, we use grid search to find the optimal softmax entropy threshold for deferring predictions to \texttt{GPT-4o}. Thresholds from 0 to 1 (in 0.01 increments) are evaluated on the validation set. At inference time, any predictions with a softmax entropy value that is above the threshold are deferred to \texttt{GPT-4o}.
\end{itemize}

\begin{table}[t!]
\centering
\small
\setlength{\tabcolsep}{6pt}
\begin{tabular}{lcc}
\toprule
\textbf{Uncertainty Metric} & \textbf{Macro F1} & \textbf{Deferral Rate} \\
\midrule
MSP & 0.81 & 0.31 \\
VET & 0.84 & 0.53 \\
MC-Dropout & 0.87 & 0.90 \\
Softmax Entropy & 0.87 & 0.90 \\
\bottomrule
\end{tabular}
\caption{Performance on Myth 1 using each deferral method. Deferral rate indicates the proportion of examples routed to \texttt{GPT-4o}.}
\label{tab:uncertainty-metrics}
\end{table}

\subsection{Results}

We apply each deferral method to the classification of Myth 1 on the gold-standard dataset to evaluate their performance. In Table~\ref{tab:uncertainty-metrics}, MC-Dropout and Softmax Entropy achieve the highest macro F1-score (0.87), but they do so by deferring 90\% of predictions to \texttt{GPT-4o}. While this boosts performance, such high deferral rates severely undermine the purpose of \methodname---effectively shifting most of the work to the expensive LLM and compromising scalability and cost-efficiency. Additionally, MC-Dropout is computationally intensive, requiring multiple forward passes per input. This further increases computational cost and latency, making it an unattractive option despite its predictive performance. 

In contrast, VET and MSP provides more practical trade-offs. VET offers strong performance (0.84 macro F1) while deferring 53\% of the predictions, and MSP offers the lowest deferral rate (31\%) with competitive performance (0.81 macro F1).  MSP offers a simple yet effective proxy for model confidence, as correct predictions tend to have higher softmax scores than incorrect ones \cite{10.5555/3295222.3295387, hendrycks2017a}. VET can complement MSP by providing class-level insights: by analyzing validation performance, we can defer predictions from classes where the model consistently underperforms (e.g., class-level F1 < 0.8), making it especially useful for handling systematic weaknesses. Thus, we prioritize MSP and VET for downstream use.

Importantly, combining MSP and VET further improves performance and coverage: in Table~\ref{tab:myth-performance-summarized}, the MSP+VET cascade achieves a macro F1-score of 0.86 while deferring only 60\% of predictions. This approach retains most of the performance gains seen in MC-Dropout and Softmax Entropy, but at a substantially lower computational and financial cost. Thus, MSP+VET strikes an effective balance between accuracy, efficiency, and scalability, making it well-suited for large-scale labeling systems.

\section{Cost Analysis}\label{appendix:cost-analysis}

As described in \S\ref{sec:applying-triage}, we applied \methodname to label 164K videos in the \texttt{OUD Recommendation Dataset} across 8 myths. Below, we compare the estimated time, financial, and environmental costs of three labeling strategies: clinical experts, \texttt{GPT-4o}, and \methodname.

\subsection{Approach 1: Clinical Expert}

As noted in \S\ref{sec:expert-annotation}, clinical experts took approximately 3 minutes per video (or ~22.5 seconds per myth) to annotate.

\begin{itemize}
[noitemsep,topsep=0pt,leftmargin=12pt]
\item\textbf{Time Cost: } Annotating the 164K-video \texttt{OUD Recommendation Dataset} would approximately take 164{,}085 \text{videos} $\times$ 3 \text{minutes per video} = \textbf{8,209.25 hours}.
\item\textbf{Financial Cost: }At the U.S. federal minimum wage (\$7.25 per hour) as the lower-bound financial cost \cite{us-min-wage}, it would cost: \$7.25 per hour * 8,209.25 hours = \textbf{\$59,517.06} for a single expert to annotate.\footnote{Under a standard annotation setup consisting of three annotators per annotation, it would require three times the time and financial costs.}
\end{itemize}

\subsection{Approach 2: \texttt{GPT-4o}}

\texttt{GPT-4o} substantially reduces time and labor costs compared to clinical experts, but remains financially and computationally expensive due to its large size (estimated 200B+ parameters \cite{gpt-4o-size}) and external API cost.

\begin{itemize}
[noitemsep,topsep=0pt,leftmargin=12pt]
\item \textbf{Time Cost:} On average, each few-shot prompt took 3.4 seconds (\S\ref{appendix:llm-evaluation}). Across 8 myths, this would total 3.4 seconds per prompt $\times$ 8 myths = 27.2 seconds per video. For 164,085 videos, it is estimated:
$164{,}085 \times 27.2$ seconds = \textbf{1,239.75 hours}.
\item \textbf{Financial Cost:} Each prompt used on average 6,066.92 input tokens and 144.01 output tokens. Using OpenAI API pricing \cite{openai_pricing}, each myth-level prompt would roughly costs (6{,}066.92 input tokens $\times \frac{\textrm{\$2.50}}{\textrm{1M input tokens}}$) + (144.01 output tokens $\times \frac{\textrm{\$10.00}}{\textrm{1M output tokens}})$ = \$0.0166. For labeling 8 myths across all 164K videos, it is estimated to cost:
$164{,}085 \times 8 \times \$0.0166$ = \textbf{\$21,790.49}.
%With the Batch API (50\% discount), the lower-bound cost is \textbf{\$10,895.25}.

\item \textbf{Environmental Cost:} Each \texttt{GPT-4o} query is estimated to consume ~3 watt-hours (Wh) \cite{DEVRIES20232191}. Total usage across all 164K videos and 8 myths would suggest:
$164{,}085 \times 8 \times 3 Wh = 3,938.04 kWh$.
At 0.374 kg CO\textsubscript{2}/kWh for the U.S. national average carbon emission \cite{epa-energy}, this estimates $3,938.04 kWh \times 0.374$ kg CO\textsubscript{2}/$kWh$ =
\textbf{1,472.83 kg CO\textsubscript{2}} in emissions.
\end{itemize}

\subsection{Approach 3: \methodname}

\methodname combines a lightweight \texttt{DeBERTA-v3-base} model with selective deferral to \texttt{GPT-4o}, reducing both financial and environmental costs while maintaining strong performance. We account for (1) training and inference of \texttt{DeBERTA}, and (2) the cost of deferring 70,777 predictions to \texttt{GPT-4o} across 8 myths in the 164K-video \texttt{OUD Recommendation Dataset}.

\begin{itemize}
[noitemsep,topsep=0pt,leftmargin=12pt]
\item \textbf{Time Cost:} Training involved 27 models per myth (3 learning rates × 3 weight decays × 3 training setups), with each model taking $\sim$60 minutes (\S\ref{appendix:distillation-details}). Across 8 myths, this is estimated to take:
$8 \times 27 \times 60$ minutes = 216 hours.

Inference on the full dataset took 16.72 hours (e.g., $\sim$ 2.09 hours per myth). 

For the 70,777 deferred examples and each few-shot prompt taking roughly 3.4 seconds (\S\ref{appendix:llm-evaluation}), \texttt{GPT-4o} is estimated to require:
$70{,}777 \times 3.4$ seconds = 66.84 hours.

Total time: 216 + 16.72 + 66.84 = \textbf{299.56 hours}.
\item\textbf{Financial Cost:} As mentioned in \S \ref{appendix:distillation-details}, \texttt{DeBERTA} training and inference (232.72 hours) was run on NVIDIA A40 GPUs (\$0.46/hr),\footnote{The rental costs of NVIDIA A40 GPUs range from \$0.40 to \$0.46 per hour on AI Cloud vendors \cite{gpu-pricing, gpu-pricing2}. We use the upper bound cost to obtain a conservative estimate of both our cost savings in comparison to other approaches.} estimated to total:
$232.72 \times \$0.46 = \$107.05$.

\texttt{GPT-4o} inference cost, which costs \$0.01666 per prompt:
$70{,}777 \times \$0.0166 = \$1,174.89$%, or \$587.45 using OpenAI's Batch API (50\% discount).

Total cost: \$107.05 + \$1,174.89 = \textbf{\$1,281.94}.
\item\textbf{Environmental Cost:} The training and inference process on NVIDIA A40 GPUs, with 300W power draw based on \citet{nvidia-power}, over 232.72 hours is estimated to consume approximately: $232.72 \times 300$ = 69.82 kWh. Additionally, deferring 70,777 instances to \texttt{GPT-4o} may consume an estimated $70,777 \times 3$ Wh = 212.33 kWh, based on prior estimates of 3Wh per prompt \cite{DEVRIES20232191}. In total, the process is estimated to use 69.82 + 212.33 = 282.15 kWh. 

At 0.374 kg CO\textsubscript{2}/kWh \cite{epa-energy}, the estimated emissions are = $282.15 \times 0.374 = \textbf{105.52 kg CO\textsubscript{2}}$.
\end{itemize}

\begin{table}[t!]
\centering
\small
\setlength{\tabcolsep}{8pt}
\begin{tabular}{lcc}
\toprule
\textbf{Myth} & \textbf{Macro F1-Score} & \textbf{Accuracy} \\
\midrule
M1 & 0.773 & 0.97 \\
M2 & 0.951 & 0.99 \\
M3 & 0.885 & 0.98 \\
M4 & 0.838 & 0.98 \\
M5 & 1.000 & 1.00 \\
M6 & 0.932 & 0.99 \\
M7 & 0.887 & 0.99 \\
M8 & 1.000 & 1.00 \\
\bottomrule
\end{tabular}
\caption{Performance of \methodname on 100 randomly-sampled videos from the \texttt{OUD Recommendation Dataset}. Ground-truth labels were obtained through manual consensus annotation by two authors following established guidelines (\S\ref{appendix:annotation-guideline}). However, note that due to class imbalance with extensive amounts of videos labeled as ``neither,'' F1-scores are highly sensitive and have high variance.}
\label{tab:myth-performance-recommendation}
\end{table}

\section{Additional Evaluation of \methodname}\label{appendix:additional-evaluation}

As an additional evaluation, we validated \methodname on 100 randomly sampled videos from the \texttt{OUD Recommendation Dataset}, following prior works \cite{Albadi_2022, dammu-etal-2024-uncultured}. Two authors independently annotated the videos using the annotation guidelines (\S \ref{appendix:annotation-guideline}), achieving a Cohen's Kappa score of 0.545---indicating ``moderate agreement'' \cite{8d20e0b8-89d8-3d65-bcf5-8c19d56ec4ab}. Then, the authors reached a consensus on labels. 

Table \ref{tab:myth-performance-recommendation} summarizes the performance across the myths.  Due to class imbalance with extensive amounts of videos labeled as ``neither'' (e.g., Myth 8 had only one ``supporting'' video and 99 ``neither'' videos), F1-scores are highly sensitive and have high variance. Nonetheless, the results indicate that \methodname reliably identified neutral or irrelevant content and showed performance comparable to results on the gold-standard dataset (\S \ref{sec:inference-triage}).

\section{Resolving Overall Stance}\label{appendix:overall-stance}

Since each video received one label per myth (8 total), we determined an overall stance label to reflect the video's holistic stance towards OUD myths. As discussed in \S \ref{sec:resolving-overall}, we manually resolved the overall stance of videos containing both \textit{supporting} and \textit{opposing} labels. This applied to 63 videos in the \texttt{OUD Search Dataset} and 193 videos in the \texttt{OUD Recommendation Dataset}.

\subsection{Manual Annotations.}\label{sec:overall-manual}
Two authors independently reviewed the 63 videos from the \texttt{OUD Search Dataset}. To determine the stance, reviewers examined \texttt{GPT-4o}’s extracted excerpts and justifications across all myths and watched each video in full. They evaluated the prominence, tone, and framing of each myth, considering how much emphasis the video placed on supporting or opposing content. Rather than simply counting myth stances, reviewers assessed the overall message. For instance, a video that opposes fewer myths may still be labeled \textit{opposing} if that content is central and persuasive. Public health implications were also considered: for example, a video that debunks a minor myth (e.g., ``Kratom is addictive'') but promotes a more harmful one (e.g., ``cold turkey is a viable method'') was labeled as \textit{supporting} OUD-related myths. 

After annotating the first 32 videos, the authors reached a Cohen’s Kappa score of 0.347, indicating fair agreement \cite{8d20e0b8-89d8-3d65-bcf5-8c19d56ec4ab}, and resolved a consensus label through discussion. After extensive discussion, they then annotated the remaining 31 videos, achieving a higher score of 0.688 (substantial agreement). Given the improved reliability and agreement, an author proceeded to annotate a random sample of 63 out of the remaining 193 videos in the \texttt{OUD Recommendation Dataset}, leaving 130 videos unannotated.

\begin{table}[t!]
\centering
\small
\setlength{\tabcolsep}{5pt}
\begin{tabular}{lccc}
\toprule
\textbf{Model} & \textbf{Accuracy} & \textbf{F1 (Macro)} & \textbf{F1 (Weighted)} \\
\midrule
GPT-4o   & 0.82 & 0.72 & 0.84 \\
\textbf{GPT-4.1}  & \textbf{0.93} & \textbf{0.79} & \textbf{0.92} \\
\bottomrule
\end{tabular}
\caption{Performance of GPT-4o and GPT-4.1 on labeling the overall stance of 126 videos related to OUD myths, evaluated against human annotations. Best performances are bolded.}
\label{tab:stance-performance}
\end{table}

\subsection{Employing LLM-as-a-judge.}

Prior work \cite{10.5555/3666122.3668142} has explored the LLM-as-a-judge paradigm as a scalable alternative to human annotation for approximating human preferences. Following this approach, and in line with other works \citet{park-etal-2024-valuescope}, we use the LLM-as-a-judge approach to assess and label the overall stance of the remaining 130 videos. Using the prompt shown in Figure \ref{appendix:overall-stance}, we evaluate the effectiveness of \texttt{GPT-4o} and \texttt{GPT-4.1} on a set of 126 human-annotated data (\S \ref{sec:overall-manual}). As shown in Table~\ref{tab:stance-performance}, \texttt{GPT-4.1} outperformed \texttt{GPT-4o}, achieving 0.93 accuracy and a macro F1-score of 0.79. Given its strong performance, we used \texttt{GPT-4.1} to scale stance annotations for the remaining 130 videos.

\section{Additional Analysis}\label{appendix:additional-analysis}

Here, we analyze the prevalence of OUD myths across search queries, topics, search filters, and compare video engagement metrics across labels.

\begin{table*}[ht]
\centering
\small
\begin{tabular}{cccc}
\toprule
\textbf{Level} & \textbf{Opposing $\rightarrow$ Supporting (\%)} & \textbf{Neutral $\rightarrow$ Supporting (\%)} & \textbf{Supporting $\rightarrow$ Supporting (\%)} \\
\midrule
Level 1 & 5.43 & 1.57 & 12.70 \\
Level 2 & 6.35 & 1.21 & 19.65 \\
Level 3 & 5.32 & 0.52 & 17.16 \\
Level 4 & 3.63 & 0.29 & 20.08 \\
Level 5 & 3.25 & 0.17 & 22.22 \\
\bottomrule
\end{tabular}
\caption{Percentage of recommended videos labeled as \textit{supporting}, broken down by the label of the source video and recommendation level (column ``Level''). For example, ``Supporting $\rightarrow$ Supporting'' at Level 1 indicates that 12.7\% of recommendations from supporting videos (e.g., source) led to another supporting video.}
\label{tab:recommendation-supporting-percent}
\end{table*}

\begin{figure}[!t]
  \centering
\includegraphics[width=0.55\linewidth]{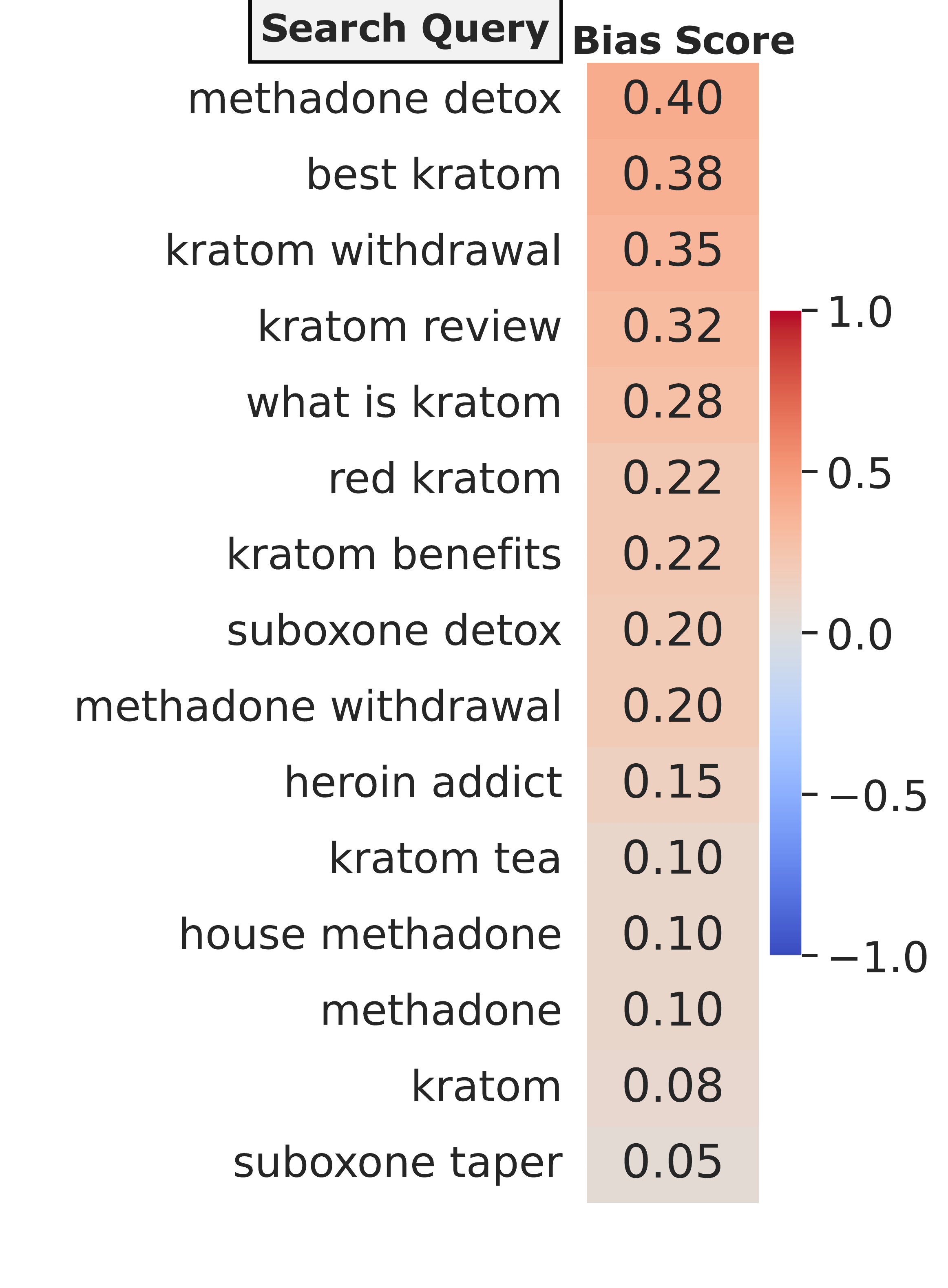}
  \caption{Top 15 search queries when sorted by the myth bias score. These queries are the most problematic ones, containing the highest amounts of myths in the search results.}
  \label{fig:query-wise-ranking}
\end{figure}

\subsection{Myth Bias Scores in Search Queries}
Figure \ref{fig:query-wise-ranking} displays the top 15 search queries with the highest myth bias scores, indicating a higher prevalence of myths. The query "methadone detox" has the highest bias score, implicitly reinforcing three myths: that methadone is as dangerous or addictive as opioids (Myth 1), that the ultimate goal of MAT is abstinence from any opioid use (Myth 3), and that detoxing is a safe and valid treatment approach (Myth 6). Notably, even seemingly neutral queries such as "methadone" and "kratom" yield search results biased toward misinformation, highlighting the pervasive influence of myth-supporting content in OUD-related search results.

\subsection{Distribution of Labels Across Myths and Topics}

\begin{figure*}[!t]
  \centering
\includegraphics[width=\linewidth]{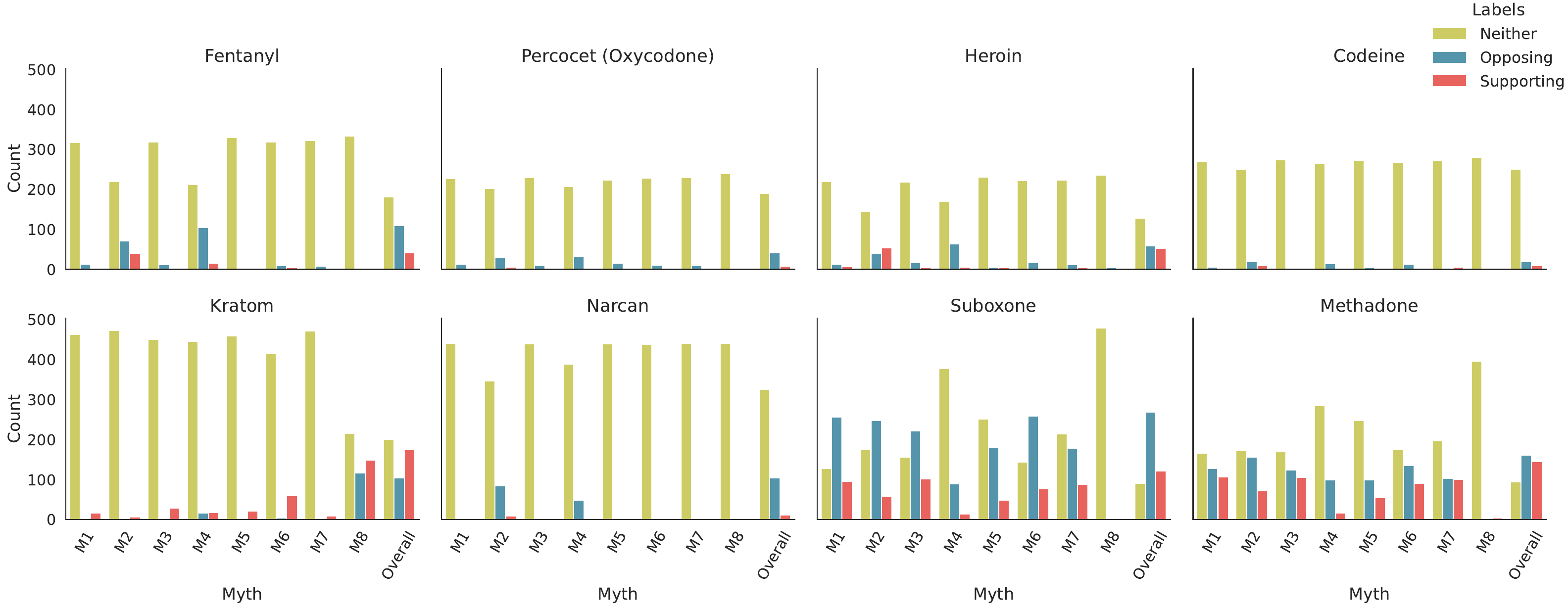}
  \caption{Frequency of labels across eight OUD-related myths and overall stance, broken down by topics. These are based on the 2.9K search results from the \texttt{OUD Search Dataset}. The subplots illustrate the variation in supporting and opposing labels across myths and topics. Topics ``Suboxone'' and ``Methadone'' consistently showed high levels of both supporting and opposing labels across myths, suggesting that they are contentious subjects in OUD-related content. Note that ``Overall'' refers to the overall stance labels, as created in \S \ref{sec:resolving-overall}.}
  \label{fig:topic-wise-myth-distribution}
\end{figure*}

Figure \ref{fig:topic-wise-myth-distribution} shows the frequency of labels across the eight myths and overall stance, grouped by topic. Methadone and Suboxone have high counts of both supporting and opposing labels, highlighting their contentious nature. In contrast, Kratom shows a high frequency of supporting labels, especially for Myth 8 (\textit{Kratom is a non-addictive and safe alternative to opioids.}).

\subsection{Distribution of Labels Across Myths and Filters}
\begin{figure*}[!t]
  \centering
\includegraphics[width=\linewidth]{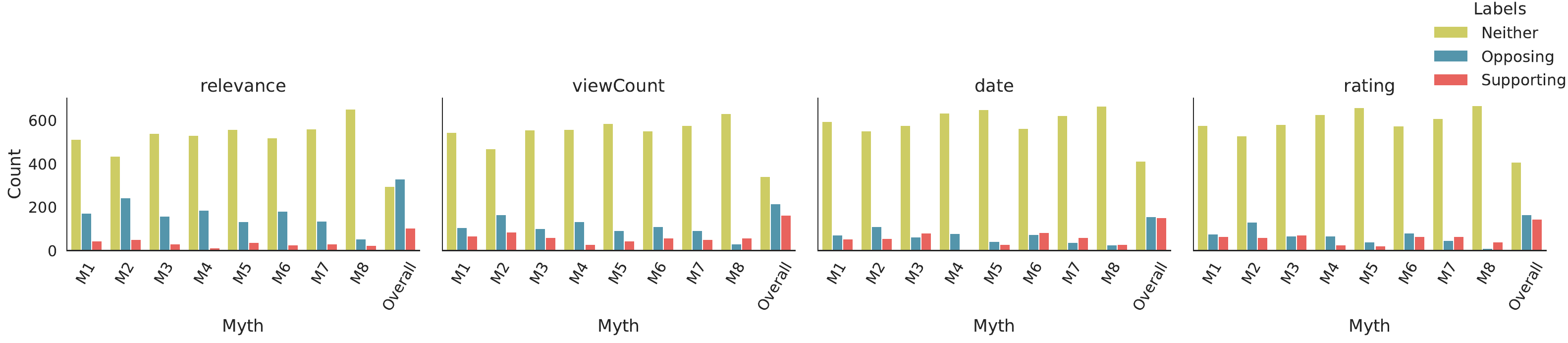}
  \caption{Frequency of labels across eight OUD-related myths and overall stance, broken down by search filters. These are based on the 2.9K search results from the \texttt{OUD Search Dataset}. The subplots illustrate the variation in supporting and opposing labels across myths and search filters.}
  \label{fig:filter-wise-myth-distribution}
\end{figure*}

Figure \ref{fig:filter-wise-myth-distribution} shows the frequency of labels across the eight myths and overall stance, grouped by filters. Sorting by relevance shows high counts of opposing labels across all myths, as discussed in \S \ref{sec:prevalence-analysis-search}. However, employing a different search filter returns relatively fewer opposing labels and more supporting labels. 

\subsection{Engagement Analysis}
A Shapiro-Wilk Test revealed that our data is not normal, so we used the non-parametric Kruskal-Wallis test to compare video engagement across stance labels (e.g., supporting, neither, and opposing). We found significant differences in views (KW H(2, N$=$1776)=10.93, \textit{p}$<$0.01, $\eta^2$=0.005) and number of comments (KW H(2, N$=$1776)=12.38, \textit{p}$<$0.01, $\eta^2$=0.006), but not in likes and like-to-view ratio. We conducted post-hoc Conover-Iman Test with Bonferroni adjustment, revealing that opposing videos tend to get more views than supporting and neither videos. Meanwhile, we found that supporting videos tend to get more comments than neither videos, indicating higher engagement and discussion around myth-supporting content.

\begin{figure*}[!t]
\centering
\fbox{\begin{minipage}{0.95\textwidth}
\footnotesize
\ttfamily
\textbf{System Persona:} You are a public health expert with specialized knowledge of opioid use disorder (OUD) and medication-assisted treatment (MAT). You are well-versed in the common myths and misconceptions about OUD and MAT that circulate on online platforms like YouTube. Your task is to analyze the provided YouTube video metadata and identify the presence of the specified OUD-related myth within the video metadata.\\

Using the provided LABEL DESCRIPTIONS, evaluate the YOUTUBE VIDEO METADATA based on its stance towards the  MYTH and justify your label with brief EXCERPT(s) from the metadata. While evaluating the video, assign one of three labels based on the video's stance towards the specified MYTH: label it as "Supports the OUD myth (1)" if it supports or promotes the MYTH, including subtle undertones and implications; "Opposes the OUD myth (-1)" if it debunks or opposes the MYTH; and "Neither (0)" if it is neither supporting nor opposing the MYTH.\\

***MYTH to evaluate starts here***\\
MYTH DEFINITION: [MYTH-DEFINITION]\\
***MYTH to evaluate ends here***\\

***LABEL DESCRIPTIONS starts here *** \\
- Supports the OUD myth (1): The video promotes or supports the OUD myth, including subtle undertones and implications. For example, a video subtly implying and promoting the provided myth in the description and transcript.\\
- Neither (0): The video neither supports nor opposes the OUD myth, including neutral information surrounding the opioid epidemic, irrelevant information that is not relevant to OUD, and unrelated information not related to the specified OUD myth. For example, news channels reporting on overdose incidents, cartoon shows, and videos that may promote other myths, but not the supporting nor opposing the specified myth.\\
- Opposes the OUD myth (-1): The video opposes, debunks, or offers countervailing contexts to the OUD myth. For example, a video opposing the provided myth in the transcript.\\
***LABEL DESCRIPTIONS ends here ***\\

Now, given what you learned from the LABEL DESCRIPTIONS above, please evaluate the YOUTUBE VIDEO METADATA on their stance towards the specified MYTH and provide justification on your label with direct, brief EXCERPT(s) from the YOUTUBE VIDEO METADATA. ONLY EXTRACT INTENTIONAL SHORT, BRIEF EXCERPTS TO JUSTIFY YOUR LABEL. Remember to assign a label based on the video’s stance: label it as "Supports the OUD myth (1)" if it supports or promotes the MYTH, including subtle undertones and implications; "Opposes the OUD myth (-1)" if it debunks or opposes the MYTH; and "Neither (0)" if it is neither supporting nor opposing the specified MYTH.\\

FORMAT your response as a JSON object in the following structure [(LABEL, EXCERPTS, JUSTIFICATION)]. Make sure to have the keys LABEL, EXCERPTS, JUSTIFICATION in the JSON structure.\\

****YOUTUBE VIDEO METADATA to be evaluated starts here ****:

- Video Title: [TITLE]\\
- Video Description: [DESCRIPTION]\\
- Video Transcript: [TRANSCRIPT]\\
- Video Tags: [TAGS]
   
****YOUTUBE VIDEO METADATA to be evaluated ends here ****.
\end{minipage}}
\caption{Zero-shot prompt used for the three-class classification task to detect an OUD-related myth in a video. In place of ``[MYTH-DEFINITION],'' we provided a specific OUD-related myths from Table \ref{tab:oud-myths}.}
\label{fig:zero-shot-prompt}
\end{figure*}

\begin{figure*}[!t]
\centering
\fbox{\begin{minipage}{0.95\textwidth}
\footnotesize
\ttfamily
\textbf{System Persona:} You are a public health expert with specialized knowledge of opioid use disorder (OUD) and medication-assisted treatment (MAT). You are well-versed in the common myths and misconceptions about OUD and MAT that circulate on online platforms like YouTube. Your task is to analyze the provided YouTube video metadata and identify the presence of the specified OUD-related myth within the video metadata.\\

Using the provided LABEL DESCRIPTIONS, evaluate the YOUTUBE VIDEO METADATA based on its stance towards the  MYTH and justify your label with brief EXCERPT(s) from the metadata. While evaluating the video, assign one of three labels based on the video's stance towards the specified MYTH: label it as "Supports the OUD myth (1)" if it supports or promotes the MYTH, including subtle undertones and implications; "Opposes the OUD myth (-1)" if it debunks or opposes the MYTH; and "Neither (0)" if it is neither supporting nor opposing the MYTH.\\

***MYTH to evaluate starts here***\\
MYTH DEFINITION: [MYTH-DEFINITION]\\
***MYTH to evaluate ends here***\\

***LABEL DESCRIPTIONS starts here *** \\
- Supports the OUD myth (1): The video promotes or supports the OUD myth, including subtle undertones and implications. For example, a video subtly implying and promoting the provided myth in the description and transcript.\\
- Neither (0): The video neither supports nor opposes the OUD myth, including neutral information surrounding the opioid epidemic, irrelevant information that is not relevant to OUD, and unrelated information not related to the specified OUD myth. For example, news channels reporting on overdose incidents, cartoon shows, and videos that may promote other myths, but not the supporting nor opposing the specified myth.\\
- Opposes the OUD myth (-1): The video opposes, debunks, or offers countervailing contexts to the OUD myth. For example, a video opposing the provided myth in the transcript.\\
***LABEL DESCRIPTIONS ends here ***\\

Below, we provide 5 EXAMPLES of the task, each example including an assigned LABEL, relevant EXCERPT(s), and justification. These examples demonstrate the evaluations of YouTube video metadata based on their stance towards the MYTH.\\
***EXAMPLES of the task starts here***\\
EXAMPLE 1 starts here ****:\\
VIDEO\_TITLE: [EXAMPLE1\_VIDEO\_TITLE]\\
VIDEO\_DESCRIPTION: [EXAMPLE1\_VIDEO\_DESCRIPTION]\\
VIDEO\_TRANSCRIPT: [EXAMPLE1\_VIDEO\_TRANSCRIPT]\\
VIDEO$\_$TAGS: [EXAMPLE1\_VIDEO\_TAGS]\\
LABEL: [EXAMPLE1\_LABEL]\\
REASONING: [EXAMPLE1\_REASONING]\\
...\\
EXAMPLE 5 starts here ****:\\
...\\
***EXAMPLES of the task ends here***\\

Now, given what you learned from the LABEL DESCRIPTIONS and the EXAMPLES above, please evaluate the YOUTUBE VIDEO METADATA on their stance towards the specified MYTH and provide justification on your label with direct, brief EXCERPT(s) from the YOUTUBE VIDEO METADATA. ONLY EXTRACT INTENTIONAL SHORT, BRIEF EXCERPTS TO JUSTIFY YOUR LABEL. Remember to assign a label based on the video’s stance: label it as "Supports the OUD myth (1)" if it supports or promotes the MYTH, including subtle undertones and implications; "Opposes the OUD myth (-1)" if it debunks or opposes the MYTH; and "Neither (0)" if it is neither supporting nor opposing the specified MYTH.\\

FORMAT your response as a JSON object in the following structure [(LABEL, EXCERPTS, JUSTIFICATION)]. Make sure to have the keys LABEL, EXCERPTS, JUSTIFICATION in the JSON structure.\\

****YOUTUBE VIDEO METADATA to be evaluated starts here ****:

- Video Title: [TITLE]\\
- Video Description: [DESCRIPTION]\\
- Video Transcript: [TRANSCRIPT]\\
- Video Tags: [TAGS]
   
****YOUTUBE VIDEO METADATA to be evaluated ends here ****.
\end{minipage}}
\caption{Few-shot prompt used for the three-class classification task to detect an OUD-related myth in a video. In place of ``[MYTH-DEFINITION],'' we provided a specific OUD-related myths from Table \ref{tab:oud-myths}. In addition, we provided 5 examples of the task, each example accompanied by a video title, description, transcript, tags, assigned labels, and reasoning for the assigned label based on the video metadata.}
\label{fig:few-shot-prompt}
\end{figure*}

\begin{table*}[ht]
\centering
\small
\setlength{\tabcolsep}{3pt}
\begin{scriptsize}
\begin{tabular}{llcccc}
\toprule
\textbf{Model} & \textbf{Prompt} & \textbf{Myth} & \textbf{Accuracy} & \textbf{F1-Score (Macro)} & \textbf{F1-Score (Weighted)} \\
\midrule
\texttt{GPT-4o-2024-08-06} & Zero-Shot & M1 & 0.830 & 0.812 & 0.824 \\
 & Zero-Shot & M2 & 0.770 & 0.705 & 0.751 \\
 & Zero-Shot & M3 & 0.813 & 0.802 & 0.810 \\
 & Zero-Shot & M4 & 0.879 & 0.628 & 0.865 \\
 & Zero-Shot & M5 & 0.787 & 0.574 & 0.741 \\
 & Zero-Shot & M6 & 0.751 & 0.717 & 0.739 \\
 & Zero-Shot & M7 & 0.852 & 0.809 & 0.847 \\
 & Zero-Shot & M8 & 0.967 & 0.818 & 0.969 \\
\cmidrule{2-6}
 & Few-Shot & M1 & \textbf{0.882} & \textbf{0.871} & \textbf{0.880} \\
 & Few-Shot & M2 & \textbf{0.849} & \textbf{0.854} & \textbf{0.850} \\
 & Few-Shot & M3 & \textbf{0.869} & \textbf{0.859} & \textbf{0.866} \\
 & Few-Shot & M4 & \textbf{0.905} & \textbf{0.818} & \textbf{0.907} \\
 & Few-Shot & M5 & \textbf{0.889} & \textbf{0.824} & \textbf{0.882} \\
 & Few-Shot & M6 & \textbf{0.869} & \textbf{0.857} & \textbf{0.866} \\
 & Few-Shot & M7 & \textbf{0.889} & \textbf{0.853} & \textbf{0.884} \\
 & Few-Shot & M8 & \textbf{0.977} & \textbf{0.866} & \textbf{0.977} \\
\midrule
\texttt{GPT-4o-mini-2024-07-18} & Zero-Shot & M1 & 0.715 & 0.691 & 0.711 \\
 & Zero-Shot & M2 & 0.708 & 0.675 & 0.701 \\
 & Zero-Shot & M3 & 0.652 & 0.633 & 0.649 \\
 & Zero-Shot & M4 & 0.741 & 0.516 & 0.756 \\
 & Zero-Shot & M5 & 0.685 & 0.586 & 0.695 \\
 & Zero-Shot & M6 & 0.646 & 0.617 & 0.640 \\
 & Zero-Shot & M7 & 0.662 & 0.623 & 0.673 \\
 & Zero-Shot & M8 & 0.728 & 0.564 & 0.810 \\
\cmidrule{2-6}
& Few-Shot & M1 & 0.816 & 0.808 & 0.816 \\
 & Few-Shot & M2 & 0.695 & 0.690 & 0.699 \\
 & Few-Shot & M3 & 0.757 & 0.752 & 0.757 \\
 & Few-Shot & M4 & 0.784 & 0.628 & 0.807 \\
 & Few-Shot & M5 & 0.711 & 0.667 & 0.734 \\
 & Few-Shot & M6 & 0.800 & 0.791 & 0.802 \\
 & Few-Shot & M7 & 0.774 & 0.747 & 0.780 \\
 & Few-Shot & M8 & 0.921 & 0.680 & 0.937 \\
\midrule
\texttt{Claude-3.5-Sonnet-20241022} & Zero-Shot & M1 & 0.790 & 0.769 & 0.784 \\
 & Zero-Shot & M2 & 0.757 & 0.720 & 0.748 \\
 & Zero-Shot & M3 & 0.761 & 0.747 & 0.754 \\
 & Zero-Shot & M4 & 0.846 & 0.587 & 0.830 \\
 & Zero-Shot & M5 & 0.731 & 0.619 & 0.726 \\
 & Zero-Shot & M6 & 0.764 & 0.737 & 0.757 \\
 & Zero-Shot & M7 & 0.790 & 0.708 & 0.775 \\
 & Zero-Shot & M8 & 0.948 & 0.688 & 0.950 \\
\cmidrule{2-6}
& Few-Shot & M1 & 0.875 & 0.864 & 0.874 \\
 & Few-Shot & M2 & 0.813 & 0.818 & 0.813 \\
 & Few-Shot & M3 & 0.846 & 0.839 & 0.845 \\
 & Few-Shot & M4 & 0.892 & 0.741 & 0.886 \\
 & Few-Shot & M5 & 0.823 & 0.743 & 0.818 \\
 & Few-Shot & M6 & 0.843 & 0.832 & 0.842 \\
 & Few-Shot & M7 & 0.839 & 0.792 & 0.832 \\
 & Few-Shot & M8 & 0.964 & 0.758 & 0.966 \\
\midrule
\texttt{Claude-3.5-Haiku-20241022} & Zero-Shot & M1 & 0.767 & 0.745 & 0.765 \\
 & Zero-Shot & M2 & 0.708 & 0.684 & 0.701 \\
 & Zero-Shot & M3 & 0.744 & 0.733 & 0.742 \\
 & Zero-Shot & M4 & 0.744 & 0.548 & 0.763 \\
 & Zero-Shot & M5 & 0.734 & 0.631 & 0.730 \\
 & Zero-Shot & M6 & 0.738 & 0.714 & 0.727 \\
 & Zero-Shot & M7 & 0.744 & 0.683 & 0.746 \\
 & Zero-Shot & M8 & 0.954 & 0.784 & 0.957 \\
\cmidrule{2-6}
& Few-Shot & M1 & 0.866 & 0.860 & 0.866 \\
 & Few-Shot & M2 & 0.721 & 0.717 & 0.718 \\
 & Few-Shot & M3 & 0.810 & 0.804 & 0.810 \\
 & Few-Shot & M4 & 0.725 & 0.578 & 0.757 \\
 & Few-Shot & M5 & 0.721 & 0.675 & 0.733 \\
 & Few-Shot & M6 & 0.813 & 0.807 & 0.814 \\
 & Few-Shot & M7 & 0.823 & 0.797 & 0.827 \\
 & Few-Shot & M8 & 0.974 & 0.860 & 0.975 \\
\bottomrule
\end{tabular}
\end{scriptsize}
\caption{Performance of \texttt{GPT-4o-2024-08-06}, \texttt{GPT-4o-mini-2024-07-18}, \texttt{Claude-3.5-Sonnet-20241022}, and \texttt{Claude-3.5-Haiku-20241022} on myth classification across different myths (M1-M8) using zero-shot and few-shot prompts. For each myth, we evaluate the performance on 305 videos from the expert-annotated gold standard dataset, excluding the five few-shot examples used in the prompt. Across all model evaluations, we found that using few-shot prompts with \texttt{GPT-4o-2024-08-06} gave the best performance across myths (\textbf{bolded}).}
\label{tab:llm-performance-part1}
\end{table*}

\begin{table*}[ht]
\centering
\small
\setlength{\tabcolsep}{3pt}
\begin{scriptsize}
\begin{tabular}{llcccc}
\toprule
\textbf{Model} & \textbf{Prompt} & \textbf{Myth} & \textbf{Accuracy} & \textbf{F1-Score (Macro)} & \textbf{F1-Score (Weighted)} \\
\midrule
\texttt{Meta-Llama-3-8B-Instruct} & Zero-Shot & M1 & 0.485 & 0.283 & 0.360 \\
 & Zero-Shot & M2 & 0.439 & 0.266 & 0.315 \\
 & Zero-Shot & M3 & 0.479 & 0.291 & 0.353 \\
 & Zero-Shot & M4 & 0.763 & 0.340 & 0.705 \\
 & Zero-Shot & M5 & 0.661 & 0.314 & 0.565 \\
 & Zero-Shot & M6 & 0.446 & 0.252 & 0.328 \\
 & Zero-Shot & M7 & 0.593 & 0.338 & 0.499 \\
 & Zero-Shot & M8 & 0.849 & 0.409 & 0.867 \\
\cmidrule{2-6}
 & Few-Shot & M1 & 0.541 & 0.509 & 0.542 \\
 & Few-Shot & M2 & 0.407 & 0.333 & 0.314 \\
 & Few-Shot & M3 & 0.554 & 0.548 & 0.559 \\
 & Few-Shot & M4 & 0.721 & 0.311 & 0.679 \\
 & Few-Shot & M5 & 0.382 & 0.318 & 0.423 \\
 & Few-Shot & M6 & 0.410 & 0.376 & 0.400 \\
 & Few-Shot & M7 & 0.638 & 0.504 & 0.612 \\
 & Few-Shot & M8 & 0.414 & 0.257 & 0.534 \\
\midrule
\texttt{Meta-Llama-3.3-70B-Instruct} & Zero-Shot & M1 & 0.761 & 0.712 & 0.744 \\
 & Zero-Shot & M2 & 0.685 & 0.573 & 0.645 \\
 & Zero-Shot & M3 & 0.675 & 0.590 & 0.636 \\
 & Zero-Shot & M4 & 0.820 & 0.492 & 0.779 \\
 & Zero-Shot & M5 & 0.741 & 0.524 & 0.700 \\
 & Zero-Shot & M6 & 0.669 & 0.593 & 0.642 \\
 & Zero-Shot & M7 & 0.715 & 0.536 & 0.667 \\
 & Zero-Shot & M8 & 0.954 & 0.766 & 0.956 \\
\cmidrule{2-6}
 & Few-Shot & M1 & 0.787 & 0.765 & 0.784 \\
 & Few-Shot & M2 & 0.767 & 0.759 & 0.767 \\
 & Few-Shot & M3 & 0.780 & 0.747 & 0.769 \\
 & Few-Shot & M4 & 0.846 & 0.630 & 0.835 \\
 & Few-Shot & M5 & 0.784 & 0.707 & 0.784 \\
 & Few-Shot & M6 & 0.810 & 0.767 & 0.798 \\
 & Few-Shot & M7 & 0.833 & 0.772 & 0.826 \\
 & Few-Shot & M8 & 0.961 & 0.762 & 0.964 \\
\midrule
\texttt{Gemini-1.5-Pro} & Zero-Shot & M1 & 0.711 & 0.674 & 0.703 \\
 & Zero-Shot & M2 & 0.672 & 0.588 & 0.645 \\
 & Zero-Shot & M3 & 0.689 & 0.655 & 0.673 \\
 & Zero-Shot & M4 & 0.797 & 0.515 & 0.785 \\
 & Zero-Shot & M5 & 0.734 & 0.574 & 0.715 \\
 & Zero-Shot & M6 & 0.702 & 0.681 & 0.688 \\
 & Zero-Shot & M7 & 0.787 & 0.724 & 0.785 \\
 & Zero-Shot & M8 & 0.931 & 0.664 & 0.938 \\
\cmidrule{2-6}
 & Few-Shot & M1 & 0.836 & 0.824 & 0.838 \\
 & Few-Shot & M2 & 0.695 & 0.692 & 0.690 \\
 & Few-Shot & M3 & 0.813 & 0.807 & 0.814 \\
 & Few-Shot & M4 & 0.803 & 0.561 & 0.781 \\
 & Few-Shot & M5 & 0.777 & 0.707 & 0.781 \\
 & Few-Shot & M6 & 0.846 & 0.830 & 0.845 \\
 & Few-Shot & M7 & 0.826 & 0.751 & 0.809 \\
 & Few-Shot & M8 & 0.951 & 0.752 & 0.955 \\
\midrule
\texttt{Gemini-1.5-Flash} & Zero-Shot & M1 & 0.718 & 0.645 & 0.687 \\
 & Zero-Shot & M2 & 0.711 & 0.578 & 0.664 \\
 & Zero-Shot & M3 & 0.705 & 0.628 & 0.664 \\
 & Zero-Shot & M4 & 0.800 & 0.394 & 0.735 \\
 & Zero-Shot & M5 & 0.715 & 0.432 & 0.639 \\
 & Zero-Shot & M6 & 0.652 & 0.592 & 0.613 \\
 & Zero-Shot & M7 & 0.741 & 0.618 & 0.704 \\
 & Zero-Shot & M8 & 0.957 & 0.754 & 0.955 \\
\cmidrule{2-6}
 & Few-Shot & M1 & 0.780 & 0.728 & 0.767 \\
 & Few-Shot & M2 & 0.721 & 0.679 & 0.711 \\
 & Few-Shot & M3 & 0.728 & 0.665 & 0.702 \\
 & Few-Shot & M4 & 0.836 & 0.588 & 0.802 \\
 & Few-Shot & M5 & 0.780 & 0.683 & 0.776 \\
 & Few-Shot & M6 & 0.820 & 0.791 & 0.814 \\
 & Few-Shot & M7 & 0.744 & 0.637 & 0.723 \\
 & Few-Shot & M8 & 0.931 & 0.712 & 0.940 \\
\bottomrule
\end{tabular}
\end{scriptsize}
\caption{Performance of \texttt{Meta-Llama-3-8B-Instruct}, \texttt{Meta-Llama-3.3-70B-Instruct}, \texttt{Gemini-1.5-Pro}, and \texttt{Gemini-1.5-Flash} on myth classification across different myths (M1-M8) using zero-shot and few-shot prompts. For each myth, we evaluate the performance on 305 videos from the expert-annotated gold standard dataset, excluding the five few-shot examples used in the prompt.}
\label{tab:llm-performance-part2}
\end{table*}

\begin{table*}[ht]
\centering
\small
\setlength{\tabcolsep}{3pt}
\begin{scriptsize}
\begin{tabular}{llcccc}
\toprule
\textbf{Model} & \textbf{Prompt} & \textbf{Myth} & \textbf{Accuracy} & \textbf{F1-Score (Macro)} & \textbf{F1-Score (Weighted)} \\
\midrule
\texttt{DeepSeek-v3} & Zero-Shot & M1 & 0.757 & 0.722 & 0.744 \\
 & Zero-Shot & M2 & 0.698 & 0.618 & 0.672 \\
 & Zero-Shot & M3 & 0.790 & 0.773 & 0.783 \\
 & Zero-Shot & M4 & 0.823 & 0.458 & 0.774 \\
 & Zero-Shot & M5 & 0.738 & 0.503 & 0.675 \\
 & Zero-Shot & M6 & 0.649 & 0.591 & 0.613 \\
 & Zero-Shot & M7 & 0.810 & 0.743 & 0.795 \\
 & Zero-Shot & M8 & 0.967 & 0.800 & 0.968 \\
\cmidrule{2-6}
 & Few-Shot & M1 & 0.852 & 0.845 & 0.851 \\
 & Few-Shot & M2 & 0.741 & 0.728 & 0.735 \\
 & Few-Shot & M3 & 0.823 & 0.809 & 0.819 \\
 & Few-Shot & M4 & 0.839 & 0.587 & 0.815 \\
 & Few-Shot & M5 & 0.820 & 0.734 & 0.811 \\
 & Few-Shot & M6 & 0.849 & 0.838 & 0.847 \\
 & Few-Shot & M7 & 0.833 & 0.766 & 0.819 \\
 & Few-Shot & M8 & 0.967 & 0.809 & 0.970 \\
 \midrule
\texttt{Qwen-2.5-72b-instruct} & Zero-Shot & M1 & 0.764 & 0.742 & 0.757 \\
 & Zero-Shot & M2 & 0.751 & 0.740 & 0.748 \\
 & Zero-Shot & M3 & 0.790 & 0.775 & 0.787 \\
 & Zero-Shot & M4 & 0.823 & 0.567 & 0.816 \\
 & Zero-Shot & M5 & 0.777 & 0.603 & 0.747 \\
 & Zero-Shot & M6 & 0.675 & 0.597 & 0.647 \\
 & Zero-Shot & M7 & 0.830 & 0.784 & 0.828 \\
 & Zero-Shot & M8 & 0.951 & 0.776 & 0.953 \\
\cmidrule{2-6}
 & Few-Shot & M1 & 0.846 & 0.829 & 0.843 \\
 & Few-Shot & M2 & 0.793 & 0.791 & 0.794 \\
 & Few-Shot & M3 & 0.807 & 0.790 & 0.801 \\
 & Few-Shot & M4 & 0.807 & 0.617 & 0.801 \\
 & Few-Shot & M5 & 0.803 & 0.716 & 0.798 \\
 & Few-Shot & M6 & 0.787 & 0.764 & 0.779 \\
 & Few-Shot & M7 & 0.859 & 0.810 & 0.853 \\
 & Few-Shot & M8 & 0.967 & 0.792 & 0.968 \\
\bottomrule
\end{tabular}
\end{scriptsize}
\caption{Performance of \texttt{DeepSeek-v3} and \texttt{Qwen-2.5-72b-instruct} on myth classification across different myths (M1-M8) using zero-shot and few-shot prompts. For each myth, we evaluate the performance on 305 videos from the expert-annotated gold standard dataset, excluding the five few-shot examples used in the prompt.}
\label{tab:llm-performance-part3}
\end{table*}

\begin{figure*}[!t]
\centering
\fbox{\begin{minipage}{0.95\textwidth}
\footnotesize
\ttfamily
\textbf{System Persona:} You are a public health expert with specialized knowledge of opioid use disorder (OUD) and medication-assisted treatment (MAT). You are well-versed in the common myths and misconceptions about OUD and MAT that circulate on online platforms like YouTube. Given the metadata of a YouTube video and prior assessments indicating whether the video supports, opposes, or is neither toward each of eight OUD-related myths, your task is to analyze the provided information and determine the video's overall stance toward OUD myths. Carefully weigh the myth-level predictions and provide a reasoned judgment: does the video overall support, oppose, or remain neither toward OUD myths?\\

You are given metadata of a YouTube videos along with prior assessments indicating the video's stance towards 8 different opioid use disorder (OUD) myths. Your task is to determine the overall stance of the video toward OUD myths based on these assessments and the provided metadata.\\

***LABEL DESCRIPTIONS starts here *** \\
- [LABEL DESCRIPTION]\\
***LABEL DESCRIPTIONS ends here ***\\

***LABELED ASSESSMENTS FOR EACH MYTH STARTS HERE*** For each myth, we provide their description, labeled assessments regarding their stance towards the myth, and select excerpts and justifications of the assessment. In some cases, such excerpts and justifications may not be provided, so please use the labels for these myths into consideration.//
MYTH 1: "Agonist therapy or medication-assisted treatment (MAT) for OUD is merely replacing one drug with another."\\
- [GPT-4o-generated labels, excerpts, and justification for Myth 1]\\
...\\
MYTH 8: "Kratom is a non-addictive and safe alternative to opioids."\\
- [GPT-4o-generated labels, excerpts, and justification for Myth 8]\\
***DESCRIPTIONS AND LABELED ASSESSMENTS FOR EACH MYTH ENDS HERE***\\

****YOUTUBE VIDEO METADATA to be evaluated starts here ****:\\
- [VIDEO METADATA]\\
****YOUTUBE VIDEO METADATA to be evaluated ends here ****.\\

***IMPORTANT GUIDELINES starts here***\\
1. Do not simply count the number of myths supported or opposed: A video may support more myths than it opposes, but still overall oppose OUD myths if the opposing content is especially prominent or central to the video's message. \\
2. Evaluate the prominence, tone, and framing of each myth: Consider how strongly the video supports or opposes each myth, and how much emphasis is given to particular myths.\\
3. Context matters: A single opposed myth that is framed clearly, prominently, and persuasively may outweigh other myth stances that are only briefly mentioned or ambiguously framed. Also, consider how these myths can help or harm public health implication. For example, even if the video negates a myth like Myth 8 (e.g., "Kratom is addictive"), but promotes a more serious one (e.g., "cold turkey is viable method"), then you should resolve it as supporting OUD myths.\\
4. Use holistic reasoning: Focus on what the video communicates overall about OUD myths, not just based on the model’s per-myth stance predictions and explanations. For example, between Myth 2 and Myth 4, think about whether the video frames the person's moral as the point of blame for them doing bad things (e.g., supporting OUD myths overall) vs. their addiction making them do bad things (e.g., opposing OUD myths overall).\\
***IMPORTANT GUIDELINES ends here***\\

Now, using what you’ve learned from the label descriptions, labeled assessments across myths, and video metadata, assign an overall stance toward OUD myths based on the label descriptions. Make sure to follow the important guidelines and provide justification on your label with direct, brief EXCERPT(s) from the YOUTUBE VIDEO METADATA and prior assessments. ONLY EXTRACT INTENTIONAL SHORT, BRIEF EXCERPTS TO JUSTIFY YOUR LABEL. Remember to assign a label based on the video’s overall stance towards OUD myths: label it as "Supports OUD myths (1)" if it supports or promotes OUD myths overall, including subtle undertones and implications; "Opposes OUD myths (-1)" if it debunks or opposes OUD myths overall; and "Neither (0)" if it is neither supporting nor opposing OUD myths overall. Be conservative with labeling "Neither (0)" given that these videos were previously assessed to be opposing and supporting some OUD myth, and thus should take these assessments into account.\\

FORMAT your response as a JSON object in the following structure [(LABEL, EXCERPTS, JUSTIFICATION)]. Make sure to have the keys LABEL, EXCERPTS, JUSTIFICATION in the JSON structure.\\
\end{minipage}}
\caption{We used an LLM-as-a-judge prompt to determine a video's overall stance on OUD-related myths when both supporting and opposing labels were present. The prompt included the same label description and video metadata as in Figure~\ref{fig:few-shot-prompt}, and filled in \texttt{GPT-4o}-generated labels, excerpts, and justifications from \methodname.}
\label{fig:overall-stance-prompt}
\end{figure*}

\end{document}